\documentclass{article}

\usepackage[english]{babel}

\usepackage[a4paper,top=2cm,bottom=2cm,left=3cm,right=3cm,marginparwidth=1.75cm]{geometry}

\usepackage{amsmath}
\usepackage{amssymb}
\usepackage[onehalfspacing]{setspace}
\usepackage{float}
\usepackage[final]{graphicx}
\usepackage[hang]{footmisc}
\usepackage{placeins}

\usepackage{comment} 

\usepackage{natbib}
\setcitestyle{authoryear,open={(},close={)}}

\title{Wordkrill: Extending Wordfish into the multidimensional political space}

\author{Benjamin Riesch}

\counterwithin{figure}{section}
\counterwithin{table}{section}
\usepackage[unicode]{hyperref}

\begin{document}
\maketitle

\begin{abstract}
Spatial models are central to the study of political conflict, yet their empirical application often depends on text-based methods. A prominent example is the Wordfish model \citep{slapin_scaling_2008}, which estimates actor positions from political texts. However, a key limitation of Wordfish is its unidimensionality, despite the well-established multidimensional nature of political competition \citep{gherghina2019dynamics}. This contribution introduces \textit{Wordkrill}, a multidimensional extension of Wordfish that retains the original model’s interpretability while allowing for efficient estimation of political positions along multiple latent dimensions. After presenting the mathematical framework of Wordkrill,  its utility through brief applications to party manifestos and parliamentary speeches is demonstrated. These examples illustrate both the practical advantages and current limitations of the approach.
\end{abstract}


\section{Introduction}

The study of political conflict has long been informed by spatial models of competition, which conceptualise political actors as being positioned along latent ideological dimensions and moving within a continuous ideological space \citep{downs1957economic, bobbio1996left}. These models assume that political preferences, party strategies, and legislative behaviour can be systematically mapped within such a framework. Moreover, research suggests that individuals naturally conceptualise political preferences in spatial terms \citep{armstrong2020analyzing}. 
Empirical applications of spatial theory increasingly rely on text-based methods to infer the ideological positions of parties and individuals, given that political discourse and text production serve as primary mechanisms through which actors articulate policy preferences and distinguish themselves from competitors \citep{wilkerson2017large, lauderdale_measuring_2016}. Compared to alternative approaches, such as expert surveys \citep{jolly2022chapel} or roll-call vote analyses \citep{hix2005power}, text-scaling methods provide a computationally efficient means of estimating preferences along latent dimensions. Furthermore, they are less susceptible to censorship or missing data, which often affect alternative estimation techniques \citep{lowe2013validating, hug2010selection}.

Text scaling methods encompass a wide range of approaches, including dictionary-based models \citep{laver2000estimating}, large language models \citep{le2024positioning}, crowdsourcing techniques \citep{benoit2016crowd}, as well as supervised \citep{laver_extracting_2003}, semi-supervised \citep{watanabe_latent_2021}, and unsupervised models \citep{jentsch2020time}. 
Most of these models require intensive preparatory work, large amounts of training data or are expensive to use. In contrast, unsupervised models simultaneously infer both the latent dimension and the positioning of documents within it \citep[293]{grimmer2013text}. Because they do not rely on training data, unsupervised approaches are particularly cost-efficient for analysing large corpora. However, identifying the underlying dimension occurs post hoc, posing the risk that factors of limited substantive interest, such as dialectal differences, may drive the most distinctive variation.

One of the most widely used unsupervised text scaling models is Wordfish \citep{slapin_scaling_2008}, which has been applied across various domains of political science research \citep{ceron2015politics, Lewandowsky2022, imre2024intra}. While Poisson-based models such as Wordfish face challenges in analysing small corpora \citep{hjorth2015computers}, are sensitive to pre-processing choices \citep{denny2018text}, entail strong assumptions on the data generation process \citep{egerod_20}, and impose simplifying assumptions about language structure due to their reliance on the bag-of-words model \citep{nanni2022political}, they often correlate strongly with the true expression of positions and approximate reality sufficiently \citep{egerod_20}. Also, they continue to serve as a foundation for more advanced methodologies \citep{diaf2023communityfish, lauderdale_measuring_2016, daubler2022scaling}.

A key limitation of Wordfish is its inability to estimate positions along multiple dimensions, despite the inherently multidimensional nature of political competition \citep{gherghina2019dynamics, bakker2012complexity, benoit2012dimensionality}. If the true number of distinguishing dimensions is misidentified -- whether underestimated or overestimated -- the inferred dimension may be misspecified, leading to biased or incomplete representations of political positioning \citep{lowe2013validating}. Expanding Wordfish to multiple dimensions is therefore a logical step toward capturing the complexity of political discourse. 
This paper introduces \textit{Wordkrill},\footnote{Krill are small, shrimp-like crustaceans that play a vital role in marine ecosystems and are found in swarms that can be incredibly dense. They are a major food source for various animals, including whales and fish. The name \textit{Wordkrill} was chosen because of the maritime theme of Wordfish and because the estimated multidimensional position clusters resemble a krill swarm.} an extension of the Wordfish model that estimates document positions in a multidimensional space. Unlike the original approach, which infers positions along a single latent dimension, Wordkrill enables the estimation of preferences across multiple dimensions without requiring prior knowledge of the underlying structure of the text. 
After introducing the model in mathematical terms, it is applied to party manifestos and parliamentary speeches to assess its performance.

\section{The underlying Wordfish model}
\label{sec:wordfish}

First, we revisit the Wordfish model \citep{slapin_scaling_2008}, which serves as the foundation for our method. 
Wordfish \citep{slapin_scaling_2008} is an unsupervised scaling algorithm to estimate the positions of documents on a latent dimension based on the frequencies or features\footnote{Often these features are words but could also take the form of $n$-grams, numbers or signs. However, since the representation based on features is more general and used in mathematical terms (e.g. document-feature-matrix), the term \textit{feature} is used in the following.} in texts.
It builds on the idea that the closeness of speakers' positions manifests itself via a similar choice of features by the speakers. 
The frequency of features is modelled via a Poisson naive Bayes generative model, whereby the model's parameters depend on the document-specific ideological position and feature-specific parameters capturing the baseline frequency and responsiveness to ideology. 
Denoting documents by $i \in D = \{1,..., I\}$ and features by $j \in V = \{1,..., J\}$, the count $\omega_{ij}$ of feature $j$ in document $i$ follows a Poisson distribution:

\begin{equation}
   \omega_{ij} \sim \mathcal{P}(\lambda_{ij})
\end{equation}
    with
\begin{equation}
    \log(\lambda_{ij})=\alpha_{i} + \psi_{j}+ \beta_{j}\theta_{i}
\end{equation}

The feature count $\omega_{ij}$ depends on a document-fixed effect $\alpha$, a feature-fixed effect $\psi$, a feature-specific weight capturing the importance of feature $j$ in discriminating between document positions $\beta$,  and $\theta$, the position of document $i$ on the latent dimension. 
To estimate the parameters, a conditional maximisation algorithm is usually diploid to iteratively compute maximum-likelihood estimates of the model's parameters. 
This follows the idea that if one assumes to know $\psi$ and $\beta$ (the feature parameters), then estimating the document parameters is just like in a Poisson regression model. Vice versa, if one assumes to know $\alpha$ and $\theta$ (the document parameters), estimating the feature parameters is also a Poisson regression model. 
This leads to an algorithm, which is iterated until convergence \citep{slapin_scaling_2008}:

\begin{enumerate}
    \item  Assume the current feature parameters are correct and fit as a Poisson regression model to estimate the document parameters.
    \item Normalize $\theta$s to mean $0$ and variance $1$ to ensure identifiability and avoid drift.
    \item Assume the current document parameters are correct and fit as a Poisson regression model to estimate the feature parameters.
\end{enumerate}

The initial values for this algorithm are obtained from descriptive measures of the feature frequencies. \citet[709]{slapin_scaling_2008} suggest taking the logged mean count of each feature as a point of departure for the feature-fixed effects $\psi$ and the logged ratio of the mean feature count of each document relative to the first document in the data frame as starting values for the document-fixed effects $\alpha$. To derive initial values for the $\beta$ and $\theta$ in the model, the starting values for the document- and feature-fixed effects are subtracted from the logged feature frequencies. Of the resulting matrix, the left and right singular vectors from the singular value decomposition are then used as starting values for $\beta$ and $\theta$.

Wordfish relies on several key assumptions: Firstly, the model builds on the bag-of-words assumption, meaning it represents documents by disregarding the order of features and solely relying on their frequency. This neglects the dependence of features, ignoring co-occurrence patterns and the syntactic or semantic interplay between features  \citep{glavavs2017unsupervised}, which can lead to a loss of meaning \citep{nanni2022political}. Evidently, this is a strong simplifying assumption, and it is easy to construct sentences or whole documents for which the order of features is crucial. However, at least at the sentence level, these cases are rare \citep[272]{grimmer2013text} and at the document level, bag-of-words approaches provide a widespread approximation to the actual meaning of the text. 
Secondly, the model assumes that features are stable in meaning \citep{jentsch2020time}, i.e. denote the same things across different contexts (e.g. topics or time). Again, counterexamples can be given effortlessly, e.g. \textit{chip} as a snack vs. short for microchip. Most wordfish applications deal with this circumstance either by narrowing down texts \citep[711]{slapin_scaling_2008}, e.g. temporally or thematically, or by assuming that the main axis of differentiation of the texts is not significantly influenced. 
Third, features that are similar in meaning (e.g. \textit{debts} and \textit{liabilities}) are assumed to be different to the same extent as any other combination of features (e.g. \textit{debts} and \textit{pizza}). Ignoring this semantic proximity makes it difficult to estimate the importance of these features. 
Fourth, the assumption of features following a Poisson distribution does not always hold: For example, the distribution of frequent features is closer to a normal distribution while infrequent features tend to be negative binomial distributed \citep{lo2016ideological}.
Finally, the model assumes that the documents differ along one main dimension -- whereby in applications to political texts it is additionally assumed that this dimension is a political one \citep{slapin_scaling_2008}. To capture the position of documents in the multidimensional political space \citep{gherghina2019dynamics}, a thematic pre-classification of texts/text sections according to the political dimensions which span this multidimensional space \citep{slapin_scaling_2008} is required. However, this preselection induces further bias, as the pre-classification by researchers allows individual convictions to flow in.  

Additionally, the model requires post-hoc interpretation of its results, meaning researchers must manually assign substantive meaning to dimensions of variation, which introduces potential bias into the analysis \citep{eshima2024keyword}.
On top of that, the model learns about the dimensionality of the data without any input and assumes a unidimensional scale. This can lead to a measured dimensionality that is either meaningless, not of interest or too simplistic. If the number of underlying dimensions is underestimated, this results in biased coefficients and misspecification of the measured dimension \citep{egerod_20}. If the dimensionality assumption holds, the measured dimension might still be meaningless, as the differences between texts stem from topics or styles rather than positions \citep{lowe2013validating, lauderdale_measuring_2016}.

Despite its limitations, Wordfish is regularly used in political science research for scaling texts. For example, \citet{ProkschSven-Oliver2009HtAP} examine German party programs and estimate the parties' positions over time. Wordfish was also applied to parliamentary debates in Italy \citep{grasso2021cabinet} and Germany \citep{Lewandowsky2022} to measure polarisation within the legislative chambers. To determine the policy positions of factions, Wordfish was also used to analyse motions at party congresses \citep{ceron2015politics}. Also, when analysing short texts, such as social media posts \citep{aydogan2019ideological}, Wordfish provides position statements of both elites and voters. 
The lively use of Wordfish is due to the straightforwardness of the method and the fact that neither annotated training material nor prior knowledge of the texts is required. The statistical proportions of the corpus alone are decisive for estimating continuous positions \citep{slapin_scaling_2008}. 
Wordfish is an efficient and language-agnostic method for analysing large amounts of political texts of different types \citep{grimmer2010bayesian, proksch2010position}. 
On top of that, the model is characterised by the simplicity of its assumptions, can be interpreted straightforwardly and adapted to the research context with reasonable ease. 

\section{Wordkrill: Expanding Wordfish to $K$ dimensions}
\label{sec:wordkrill}

To consider the complexity of the space in which the speakers position themselves, the Wordfish model is extended to $K \in \mathbb{N}$ dimensions: the Wordkrill model.\footnote{The model and the functions presented are available as an R package at \url{https://github.com/naiveError/wordkrill}. The code is based on the \texttt{austin} package \citep{Austin_2017} and has been adapted accordingly.} 
Again, denoting documents by $i \in D = \{1,..., I\}$ and features by $j\in V = \{1,.., J\}$, the count of feature $j$ in document $i$ follows a Poisson distribution:
    \begin{equation}
        w_{ij} \sim \mathcal{P}(\lambda_{ij})
    \end{equation}
    but with
    \begin{equation}
        \log(\lambda_{ij})=\alpha_{i} + \psi_{j}+ \beta_{j}^{(1)}\theta_{i}^{(1)} + ...+\beta_{j}^{(K)}\theta_{i}^{(K)}
    \end{equation}
Just as in the Wordfish model, the feature count $w_{ij}$ depends on a set of document-fixed effects $\alpha$,  feature-fixed effects $\psi$, feature-specific weights $\beta_{j}^{(1)}, ..., \beta_{j}^{(K)}$ capturing the importance of feature $j$ in discriminating between document positions,  and $\theta_{i}^{(1)},...,\theta_{i}^{(K)}$ the positions of document $i$ on the $K$ latent dimensions.\footnote{Note that the superscript numbers in brackets are not powers but superindices.} 
Again, a conditional maximum likelihood approach estimates the model's parameters. Supposing to know the feature parameters, we get a Poisson regression model to estimate the document parameters and vice versa.
With some initial values, we iteratively converge to the optimal solution of our problem. 

To ensure identifiability, a zero mean and a variance of one of the documents' positions on all $K$ dimensions is assumed. In addition, it is required to control the covariance of the position estimates; setting them to zero is a convenient choice \citep{lowe2016scaling}. This is done to avoid overlapping, drift and interference. 
A nonlinear optimisation with constraints is used instead of standardising and orthogonalising the values in each iteration step. Only those solutions are permitted for which the conditions are in $\epsilon$ proximity to the respective target values.\footnote{No strict constraints are used, as the empirical estimates of the constrained variables may deviate from the true parameters by random chance.} This results in continuous, admissible solution sets on which optimisation can be performed, and the empirical uncertainty of the mathematical measures is taken into account. 
I suggest estimating the parameters jointly instead of iteratively. 

To derive initial values, the proposed approach by \citet[709]{slapin_scaling_2008} is generalised to $K$ dimensions. For the feature-fixed effects ($\psi$), the logged mean counts of each feature are chosen as the initial values. For the document-fixed effects ($\alpha$), the logged ratio of the mean feature count of each document relative to the first document in the corpus is used as starting values. The initial values for $\beta$ and $\theta$ again build on the difference of the logged feature frequencies and the document- and feature-fixed effects. Of the resulting matrix the first $K$ left and right singular vectors are then used as starting values for $\beta^{(1)},..., \beta^{(K)}$ and $\theta^{(1)},..., \theta^{(K)}$. To ensure that the initial $\theta$ values are accepted by the optimisation function, they are standardised and orthogonalised.\footnote{If no initial values are chosen, one may draw them from a standard normal distribution for all coefficients. To ensure acceptance by the optimisation function, the $\theta$ values should be standardised and orthogonalised. Note that these initial values are crucial for the estimation algorithm, and carelessly chosen initial values may bias the result.}

The proposed model inherits the disadvantages of the Wordfish model listed above, but removes the dimensionality restriction. In addition, the constraints to ensure identifiability and avoid drift mean that the estimated dimensions are empirically uncorrelated. For applications in which the dominant differences are assumed to be dependent, this can lead to distortion of the position space.

\paragraph{Select appropriate $\epsilon$} \hspace{5mm}

Which $\epsilon$ is appropriate to constrain the mean, covariance, and variance estimates depends on the application and the sample size. Since we do not have precise distribution assumptions for the position parameters, I suggest two options: firstly, the sole orientation towards the distribution of the mean and secondly, the introduction of a normal distribution assumption. For the first option, the central limit theorem gives us

\begin{equation}
    \overline{\theta} \sim \mathcal{N}(0 , \frac{1}{n})
\end{equation}

with number of cases $n$ and assuming $\mathbb{E}[\theta]=0$.\footnote{The Central Limit Theorem states that the distribution of sample means converges to a normal distribution with rising $n$, regardless of the population's distribution.} We want to select an $\epsilon$ neighbourhood in such a way that the empirical mean of a variable distributed with an expected value of $0$ lies within it with a probability of $1-\alpha$. Consequently, we allow all solution sets for which the mean values of the position estimates lie in this neighbourhood.  
Depending on a chosen level $\alpha$, we choose the value $\epsilon$ so that 

\begin{equation}
    \mathbb{P}(|\overline{X}|\leq \epsilon) \geq 1-\alpha
\end{equation}

which leads to  

\begin{equation}
    \epsilon_{\overline{X}} = z_{1-\alpha/2}\frac{1}{\sqrt{n}}
\end{equation}

where $z_{1-\alpha/2}$ is the $1-\alpha/2$ quantile of the standard normal distribution. 
Alternatively, we can introduce a normal distribution assumption of the position parameters. 

\begin{equation}
    \theta^{(1)}, ...,\theta^{(K)} \overset{i.i.d.}{\sim} \mathcal{N}(0,1)
\end{equation}

This is, in general, a reasonable assumption and allows for computation of standard errors, construction of confidence intervals and hypothesis testing \citep{slapin_scaling_2008}. It also corresponds to an intuition of political positions and is also assumed in related models \citep{lauderdale_measuring_2016}. 
The distribution of the empirical variance is then:

\begin{equation}
    \frac{(n-1)S^2}{\sigma^2} \overset{\sigma^2=1}{=} (n-1)S^2 \sim \chi^2_{n-1}
\end{equation}

Following the line of reasoning from above, we get: 

\begin{equation}
    \epsilon_{S^2} = \max\left(\left| \frac{n-1}{\chi^2_{\alpha/2, n-1}} -1 \right|,\left|\frac{n-1}{\chi^2_{1-\alpha/2, n-1}} -1\right|\right)
\end{equation}

Under the normality assumption, we also obtain suitable values for the assumed independent position parameters. The covariance, as the sum of normally distributed variables, again follows a normal distribution.

\begin{equation}
    \hat{Cov}(\theta^{(k)}, \theta^{(l)}) \sim \mathcal{N}\left(0, \frac{1}{n-1}\right)
\end{equation}

Consequently and following the considerations on $\epsilon$ neighbourhoods, we choose for $\epsilon$:

\begin{equation}
    \epsilon_{\hat{Cov}} = z_{1-\alpha/2}\frac{1}{\sqrt{n-1}}
\end{equation}

Note that this critical value is always greater than the value specified via the mean value. To finally ensure that all target values are in a $\epsilon$ proximity around their assumed values, we thus choose $\epsilon$ as follows:

\begin{equation}
    \epsilon = \max\left(\epsilon_{S^2}, \epsilon_{\hat{Cov}}\right) = \max\left(\left| \frac{n-1}{\chi^2_{\alpha/2, n-1}} -1 \right|,\left|\frac{n-1}{\chi^2_{1-\alpha/2, n-1}} -1\right|, z_{1-\alpha/2}\frac{1}{\sqrt{n-1}}\right)
\end{equation}

\paragraph{Estimating standard errors in the Wordkrill model} \hspace{5mm}

For the Wordfish model, several approaches have been proposed to estimate confidence intervals around the estimated position parameters. One common approach is based on the Fisher information matrix \citep{slapin_scaling_2008, lowe2013validating}. 
This method relies on the assumptions that the model is correctly specified, that there is sufficient data for the curvature of the log likelihood to be approximately quadratic and a proper estimation of the parameters to assume asymptotic normality \citep{lowe2013validating}. In the one-dimensional case, the observed Fisher information matrix \( I(\hat{\theta}) \) is given by the negative of the Hessian matrix of the log-likelihood function:

\begin{equation}
    I(\hat{\theta}) = - \frac{\partial^2 \ell(\theta)}{\partial \theta \partial \theta^\top} \Bigg|_{\theta = \hat{\theta}}
\end{equation}

This matrix serves as an approximation of the variance-covariance matrix of the estimator:

\begin{equation}
    \text{Var}(\hat{\theta}) \approx I(\hat{\theta})^{-1}
\end{equation}

The standard errors are then obtained from the square roots of the diagonal elements of the inverse Fisher information matrix:

\begin{equation}
    \text{SE}(\hat{\theta}_i) = \sqrt{ \left[ I(\hat{\theta})^{-1} \right]_{ii} }
\end{equation}

By drawing samples from a multivariate normal distribution with mean $ \hat{\theta}$ and covariance matrix $I(\hat{\theta})^{-1}$, simulated parameter values can be generated for uncertainty quantification \citep{slapin_scaling_2008}. 
This approach relies on interpreting the log-likelihood in the Wordfish model as an approximation to a multinomial model of feature counts, conditional on document length. Under this assumption, the Poisson likelihood and the multinomial likelihood become asymptotically equivalent \citep{lowe2011scaling},\footnote{This equivalence arises because fixing the document length makes the Poisson likelihood factorize similarly to the multinomial, with feature probabilities as functions of the latent position $\theta$.)} and the use of the observed information (i.e., the negative Hessian) to estimate standard errors remains valid.

The standard error estimation based on the observed Fisher information matrix can be naturally extended to a $K$-dimensional formulation of the Wordfish model. In this multivariate case, each document $i$ is assigned a position vector $\boldsymbol{\theta}_i \in \mathbb{R}^K$, representing positions along $K$ latent dimensions. 
The log-likelihood function now depends on the entire vector $\boldsymbol{\theta}_i$, and the observed Fisher information matrix becomes the negative Hessian matrix of second derivatives:

\begin{equation}
    I(\hat{\boldsymbol{\theta}}_i) = - \frac{\partial^2 \ell(\boldsymbol{\theta}_i)}{\partial \boldsymbol{\theta}_i \, \partial \boldsymbol{\theta}_i^\top} \Bigg|_{\boldsymbol{\theta}_i = \hat{\boldsymbol{\theta}}_i}
\end{equation}

This matrix provides an estimate of the variance-covariance matrix of the multivariate maximum likelihood estimate:

\begin{equation}
    \text{Cov}(\hat{\boldsymbol{\theta}}_i) \approx I(\hat{\boldsymbol{\theta}}_i)^{-1}
\end{equation}

The standard errors for each dimension \( k = 1, \dots, K \) are then obtained from the square roots of the diagonal elements of the inverse Fisher information matrix:

\begin{equation}
    \text{SE}(\hat{\theta}_{i}^{(k)}) = \sqrt{ \left[ I(\hat{\boldsymbol{\theta}}_i)^{-1} \right]_{kk} }
\end{equation}

As in the unidimensional case, simulation-based inference can be performed by drawing samples from a multivariate normal distribution with mean $\hat{\boldsymbol{\theta}}_i$ and covariance matrix $ I(\hat{\boldsymbol{\theta}}_i)^{-1}$. This allows uncertainty quantification for downstream tasks such as clustering, visualisation, or hypothesis testing in latent space.

As an alternative method for estimating confidence intervals of the Wordfish estimators, \citet{slapin_scaling_2008} propose a parametric bootstrap procedure \citep{davison1997bootstrap}. This approach is preferable when the model is reasonably well specified, but there is uncertainty regarding the robustness of the estimators, particularly in the case of small sample sizes \citep{lowe2013validating}. 
In this procedure, the estimated model parameters are used to calculate the expected feature counts  $\omega_{ij}$ for all $i \in D = \{1,..., I\}$ and $j \in V = \{1,..., J\}$. Based on these values, \citet{slapin_scaling_2008} suggest generating a large number (e.g., 500) of new document-feature matrices by drawing feature counts from a Poisson distribution with mean $\omega_{ij}$ for all $i \in D$ and $j \in V$. 
For each of these simulated matrices, the Wordfish algorithm is re-estimated using the original parameter estimates as starting values. The resulting distribution of parameter estimates can then be used to construct confidence intervals by extracting the upper and lower $\alpha$-quantiles. 
This bootstrap-based approach can be directly applied to the Wordkrill model introduced here.
The main drawback of this method is its computational cost: re-estimating the model hundreds of times can be very time-consuming, particularly for large datasets. In such cases, one may reasonably rely on the standard asymptotic approach for estimating confidence intervals.

\section{Empirical Application}
\label{sec:empirical_application}

Whether the proposed model measures political preference in a multidimensional political space and is suitable for the usual areas of application needs to be clarified in a comprehensive validation procedure. 
While automated text analysis reduces the time and costs of analysing large corpora, there is no guarantee that the model's output is meaningful or useful \citep{grimmer2013text}. As illustrated in the previous section, the Wordkrill model carries a lot of unrealistic assumptions regarding the language-generating process. Although aware of these drawbacks, it is unclear what consequences they have for the quality of estimates, especially since the properties to be estimated are unobservable \citep{lowe2013validating}. 
The validation of unsupervised models is challenging, and it is recommended to use a combination of experimental, substantive, and statistical evidence to show equivalence to supervised models \citep{budge2007they, slapin_scaling_2008, adcock2001measurement}. 

Previous applications show that the performance of the underlying Wordfish model is highly dependent on the application context \citep[293]{grimmer2013text}. Initial validation steps are therefore undertaken here, focusing primarily on applications of the algorithm to common text types in political science research. It is analysed whether the dimensions found are meaningful and match the expected ideas about the positioning of the documents.
More specifically, this short analysis examines German party manifestos, as well as parliamentary speeches from the Irish Dáil and the German \textit{Bundestag}.

All documents were transformed into document-feature matrices (DFM), which contain the frequency of features in the examined documents. 
This transformation to a DFM usually involves several preprocessing steps depending on the type of analysed document and the language in which the document is written \citep{denny2018text}. Following suggestions for the Wordfish approach,\footnote{There are no clear indications as to which preprocessing steps should be taken for unsupervised models \citep{slapin_scaling_2008, grimmer2010bayesian}. Thus, the \texttt{preText} package \citep{preText_2018}, which provides data-based recommendations on which preprocessing steps should be taken, is used to forge robustness checks. The method takes a subsample of the corpus under investigation and applies various combinations of the preprocessing steps \textit{using $n$-grams}, \textit{stemming}, \textit{removing stopwords}, \textit{removing punctuation}, \textit{removing numbers}, \textit{removing infrequent terms} and \textit{lowercasing}. Based on the relative distance of documents -- depending on the used preprocessing techniques --, statements can then be made about how robust the results of qualitative methods are and the choice of preprocessing steps can be justified \citep{denny2018text}.} preprocessing of the texts includes lematisation,\footnote{Lemmatisation refers to reducing features to their basic form, the so-called lemma \citep{manning2008information}. An alternative to this is stemming, which reduces the features to their stems. For some languages such as German, this is not advised as it reduces the effectiveness of bag-of-words approaches  \citep{bruinsma2019validating}.} lowercasing all features,\footnote{This can lead to a loss of information in for example in German texts, as some verbs and nouns can only be distinguished by the capital letter (e.g. \textit{bauten} -- past tense of to build; \textit{Bauten} -- buildings).} and removing punctuation, numbers, infrequent terms, and stop words \citep{grasso2021cabinet, ProkschSven-Oliver2009HtAP, slapin_scaling_2008}. 
Instead of analysing the occurrence of $n$-grams \citep{manning1999foundations}, unigrams are usually assumed to be sufficient for political text \citep{grimmer2013text,  hopkins2010method}.\footnote{Analysing $n$-grams instead of individual features leads to an exponential vocabulary growth in $n$. This, in turn, poses computational challenges, but also increases the demands on the amount of data fed in, since all combinations of tokens must occur acceptably often in the text corpus \citep{welbers2017text}. Therefore, the general go-to is analysing unigrams and adding context-dependent $n$-grams, such as \textit{White House} or \textit{Row v. Wade}, where analysing unigrams alone would cause a severe loss of information \citep{jurafsky2009speech}.} 

\subsection{German party manifestos}
\label{subsec:german_manifestos}

The first test case for the proposed method is an analysis of German party manifestos, a common benchmark also used in conjunction with the Wordfish model \citep{slapin_scaling_2008}. For this purpose, party platforms from the Manifesto Project Dataset [MPD] \citep{Lehmann2024CMP} for the years 2013, 2017, and 2021 are analysed.\footnote{The 2013 election manifesto of the \textit{Alternative für Deutschland} (AfD) was excluded due to its brevity.} 
During this period, the German party landscape remained relatively stable, and the selected years are close enough in time to largely rule out a strong temporal effect. The manifestos are analysed using a two-dimensional Wordkrill model, reflecting that the German political space primarily spans two ideological dimensions -- a socio-economic and a socio-cultural dimension \citep{grande_civil_2023, brauninger_party_2019}.\footnote{For control reasons, a three-dimensional Wordkrill model was also calculated. It turns out that the additional dimension results primarily from time differences in the election programmes and that no distinction in terms of content is relevant here.}

\begin{figure}[ht]
    \centering
    \includegraphics[width=1\textwidth]{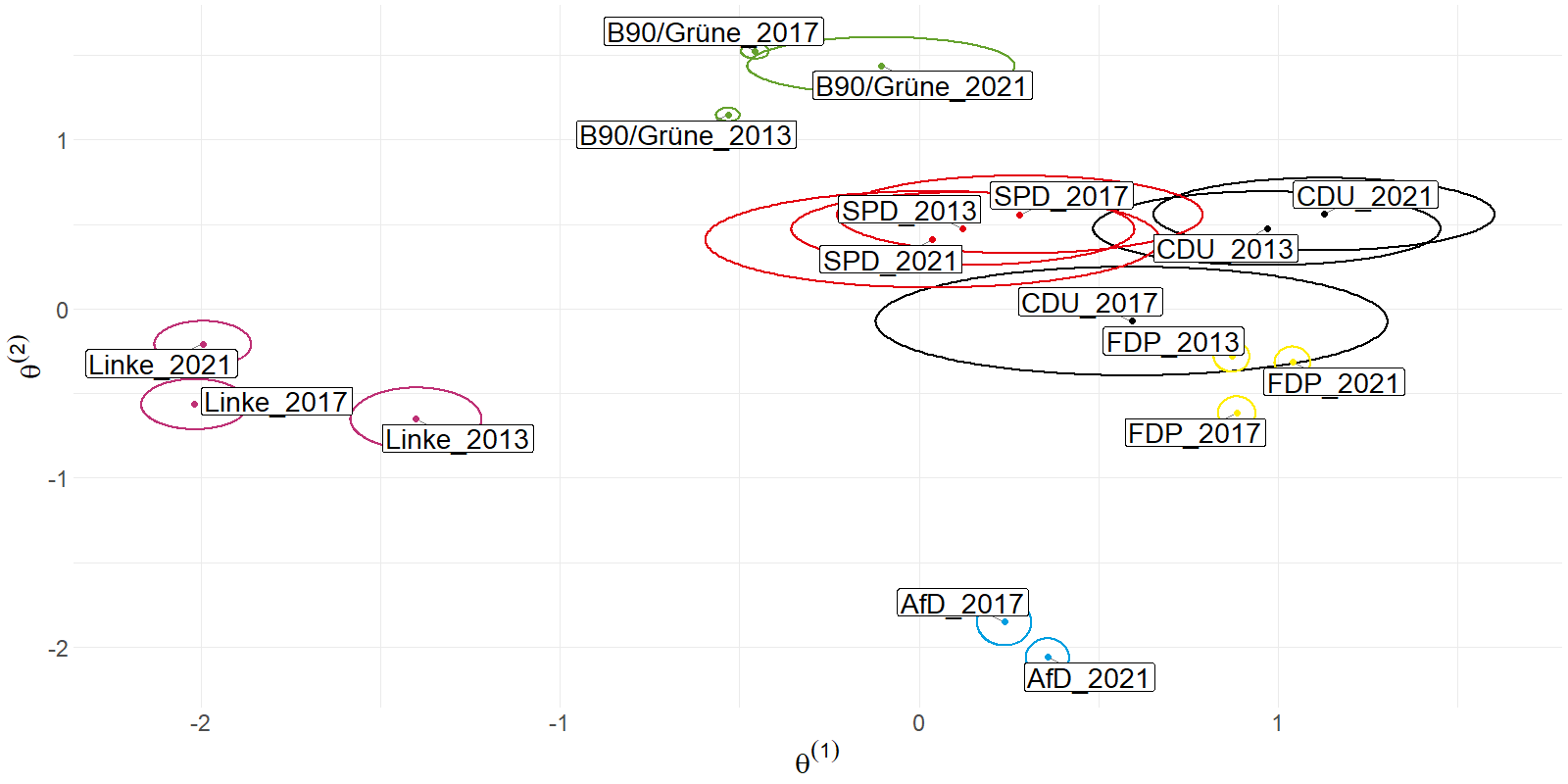}
    \caption{Position of German party manifestos based on a two-dimensional Wordkrill model with asymptotic confidence ellipses}
    \label{fig:scatter_GER_wk2}
\end{figure}

Figure \ref{fig:scatter_GER_wk2} presents the estimated positions of parties along the two dimensions. The results show that, over time, the positions of the six analysed parties remain relatively consistent. Parties that formed governing coalitions during the observed period -- such as the Greens, FDP, SPD, and CDU -- cluster closely together. In contrast, the far-right \textit{Alternative für Deutschland} (AfD) and the left-wing \textit{Die Linke} appear more isolated, each occupying extreme positions along one of the two axes.

The substantive meaning of the two dimensions can only be interpreted post hoc, whereby $\beta$-$\psi$ plots provide information. Features with high estimated $\beta$ values are especially informative for the respective positional dimension, while $\psi$ reflects the overall frequency of a feature across all documents. 
Figure \ref{fig:scatter_GER_wk2_beta1} in the appendix displays such a plot for the first dimension. The most influential features suggest a classic economic left–right dimension, with terms such as \textit{Anstrengung} (effort), \textit{Wagniskapital} (venture capital), and \textit{wettbewerbsfähig} (competitive) appearing on one end of the spectrum, and terms like \textit{Gerechtigkeit} (justice), \textit{Ausbeutung} (exploitation), and \textit{Löhne} (wages) on the other. 
Some influential terms, however, originate from adjacent issue areas -- such as \textit{Waffenexporte} (arms exports) or \textit{Nahverkehr} (local transport) -- or are only interpretable within their textual context (e.g., \textit{vgl.} [cf.] or \textit{Grundordnung} [basic order]). The socio-economic dimension appears to overlap strongly with the measured dimension, but neighbouring topics slightly distort the estimates. 

Figure \ref{fig:scatter_GER_wk2_beta2} in the appendix is consulted to interpret the second dimension. This dimension corresponds to the GAL–TAN axis (Green–Alternative–Libertarian vs. Traditional-Authoritarian-Nationalist). Negative values are associated with terms like \textit{solidarisch} (solidary), \textit{ökologisch} (ecological), and UN. In contrast, positive values reflect features such as \textit{Muslime} (muslims), \textit{Ausländer} (foreigners), or \textit{Land} (homeland). 
As with the first dimension, terms from adjacent domains (e.g., \textit{Bundesbank} [Federal Bank], \textit{Grunderwerbssteuer} [real estate transfer tax]) or terms with unclear semantic relation to other features (e.g., \textit{kündigen} [to terminate], \textit{verbleiben} [to remain]) also appear. This dimension coincides less clearly with the socio-cultural dimension, which makes interpretation more difficult. 

\begin{figure}[ht]
    \centering
    \includegraphics[width=1\textwidth]{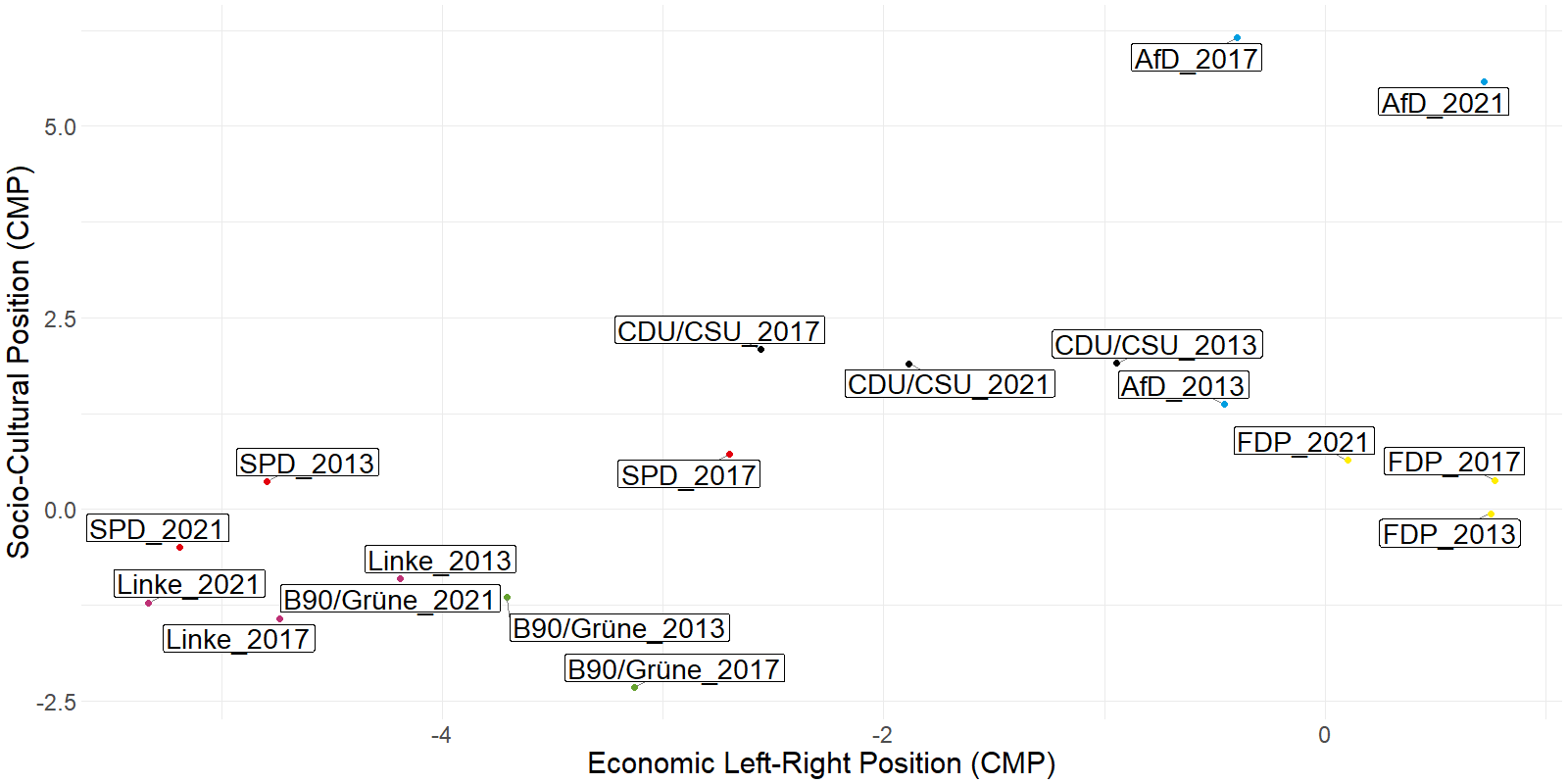}
    \caption{Position of the German parties on the socio-cultural and socio-economic dimensions (Data from the MPD, after \citet{ferreira_da_silva_three_2023})}
    \label{fig:scatter_GER_CMP_SESC}
\end{figure}

Concerning the significance of the dimensions, there are inconsistencies in the estimates, e.g. concerning the socio-cultural positions of the Left and FDP, which are estimated to be more conservative than those of the CDU. To better assess the validity of the estimates, consider Figure \ref{fig:scatter_GER_CMP_SESC}. Here, the measurements of the socio-economic and socio-cultural axes proposed by \citet{ferreira_da_silva_three_2023} are plotted based on the MPD codings.\footnote{\citet{ferreira_da_silva_three_2023} estimate the socio-economic position based on attitudes towards the economy and the welfare state. The socio-cultural dimension includes coding from the areas of nationalism, traditionalism, economic growth, multiculturalism and law \& order.} 
It is striking that the general orientation of the axes issued by Wordkrill is correct -- the Left and FDP span the socio-economic dimension, while the Greens and AfD span the socio-cultural dimension. Along the socio-economic axis, however, the distances deviate strongly from the MPD-based measures (e.g. relative distance of the Greens to the Left and the FDP). The discriminatory power between parties is also less pronounced than expected in some areas (e.g. positions of the CDU and FDP). 
In addition, the AfD's positioning is more left-wing than the coding of its manifestos suggests. 
Similar anomalies and unexpected patterns can be recognised along the identified socio-cultural axis. For example, the CDU is rated significantly more liberal by Wordkrill than by codings. Apart from this, the order and relative distance are comparable with the coding results.\footnote{Comparable results are provided by the comparison with expert opinions, see Figure \ref{fig:scatter_GER_CHES_SESC} in the appendix. Here, the opinions measured in the \textit{Chapel Hill Expert Survey} \citep{jolly2022chapel} are plotted. The measurement dates do not coincide perfectly with the German election years, but are reasonably close in time. It should also be noted that the experts' assessment is not based on the party programmes, which can lead to deviations. The incorrect estimation of the party positions by Wordkrill can also be partly explained by the salience of the dimensions in the parties, see Figure \ref{fig:scatter_GER_CHES_SESC_SAL} in the appendix. Only if topics are salient within the parties are they reflected in the party programmes. For example, the minor importance of the socio-economic dimension could lead to the AfD being incorrectly mapped on this dimension.} 
This connection may also result from the nature of party manifestos: These are written to give the party as much of an advantage as possible in the upcoming election -- with the party's history, the behaviour of members and questions of credibility limiting the content of the programme \citep{ecker_how_2022}. Overall, this distorts the \textit{true} position of the parties, which can rather be estimated by expert opinion. In addition, the coding of party programmes reveals moderation errors, whereby extreme party positions are shifted towards the centre and their true position is underestimated \citep{mikhaylov2012coder}. Against this background, the Wordkrill estimation is a method that works closer to the data and thus reflects truer self-representations of the parties' positions.

\subsection{Speeches in the Irish Dáil}
\label{subsec:IRL_parliament}

As a second example of application, speeches from the Irish Dáil (lower house of the Irish Parliament) are analysed using the proposed method. More precisely, speeches from 2009 in the debate following the presentation of the Irish budget for 2010 are analysed.\footnote{These speeches are a well-known application example for Wordfish \citep{lowe2013validating} included in the quanteda package \citep{quanteda2018} and therefore an exciting test for Wordkrill. With feature counts between 900 and nearly 8000, these speeches are of medium length and vary considerably in length.} In 14 speeches by the 5 parties represented in the Dáil\footnote{Fianna Fáil (FF), Fine Gael (FG), Labour Party (LAB), Sinn Féin (SF) and the Green Party (Green).} the speakers debated one of the most controversial budgets in Irish history. Rifts ran through the party landscape with those in favour of the cuts as necessary measures on the one side and those critical of the budget and the government on the other. On the side of the critics of the draft budget were leading members of the Labour Party and Fine Gael, as well as the anti-system party Sinn Féin. 
Taoiseach (Prime Minister) Brian Cowen and Finance Minister Brian Lenihan of the ruling Fianna Fáil were speaking on the side in favour. The three green ministers, Gormley, Cuffe, and Ryan, although committed to the budget through their coalition agreement, expressed concerns. 
In the application Wordfish, \citet{lowe2013validating} assume that the speakers' positions differ primarily on one dimension between fiscal austerity on the one hand and social protection on the other.

\begin{figure}[ht]
    \centering
    \includegraphics[width=1\textwidth]{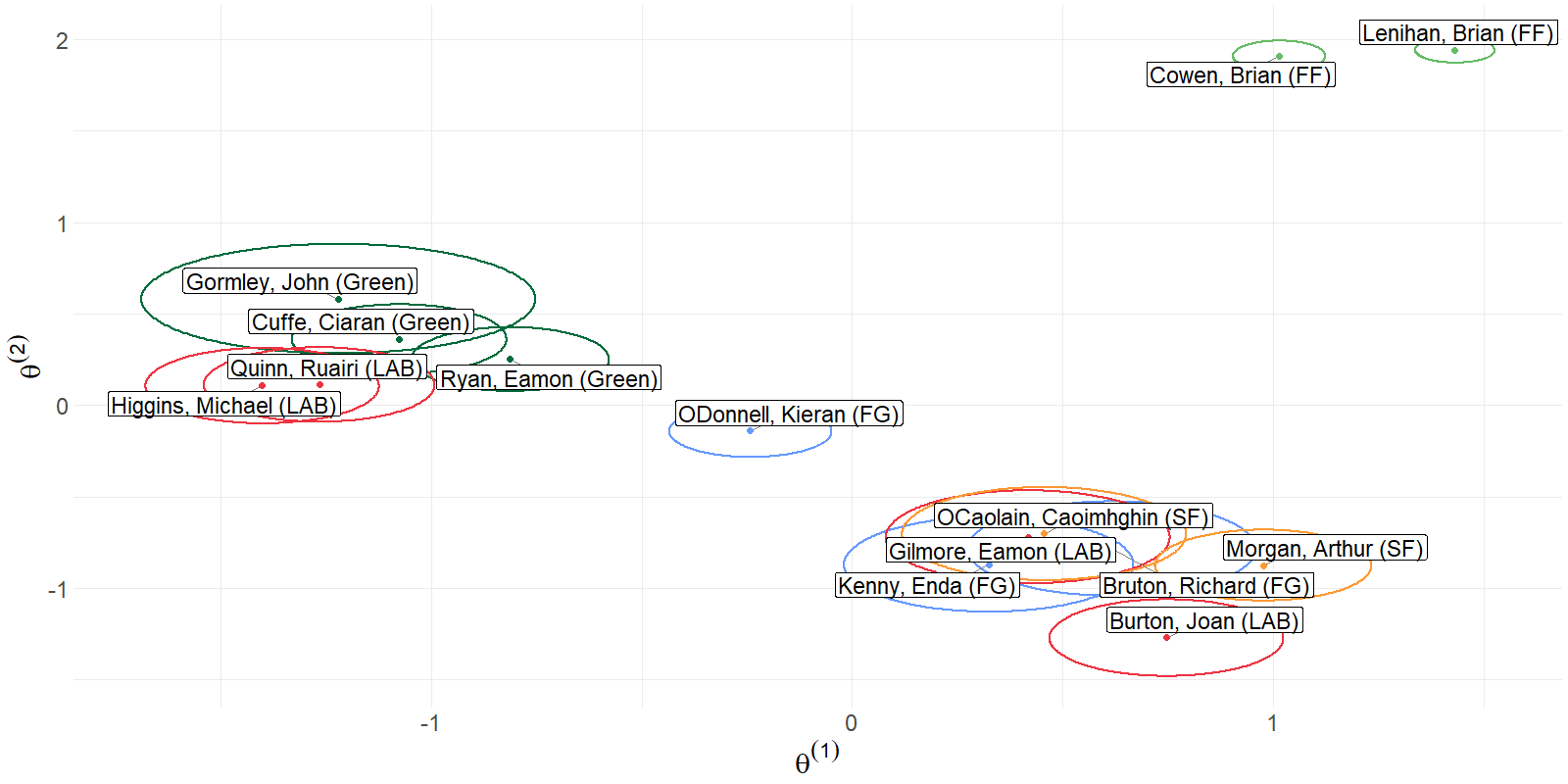}
    \caption{Position of Irish party members on two dimensions based on speeches in the Irish Dáil}
    \label{fig:scatter_IRL_wk2}
\end{figure}

The speakers' positions estimated by Wordkrill in two dimensions are displayed in Figure \ref{fig:scatter_IRL_wk2}. Three distinct clusters emerge: Fianna Fáil supporters of the budget occupy the upper-right quadrant; Green Party and some Labour MPs cluster on the left; and a third group—comprising Fine Gael, Sinn Féin, and other Labour MPs—appears in the lower-right quadrant. Fine Gael’s Kieran O'Donnell stands apart from these groupings. To interpret the underlying dimensions, the $\beta$–$\psi$ plots in Figures \ref{fig:scatter_IRL_wk2_beta1} and \ref{fig:scatter_IRL_wk2_beta2} provide insight. The first dimension captures tensions between economic pragmatism and social concern, with terms such as \textit{social}, \textit{welfare}, \textit{reduction}, and \textit{spending} at one end, and \textit{people}, \textit{society}, \textit{healthier}, and \textit{sick} at the other. Bridging terms like \textit{business}, \textit{future}, and \textit{stimulus} suggest overlap. The second dimension contrasts partisan rhetoric with technocratic discourse: negative values are marked by terms like \textit{party}, \textit{minister}, \textit{Fianna}, and \textit{alternative}, indicating politicised framing, whereas positive values highlight \textit{investment}, \textit{scheme}, \textit{efficiency}, and \textit{innovation}, pointing to a policy-focused and managerial pole. Bridging terms such as \textit{government} and \textit{budget} suggest an intermediary role between political authority and policy implementation.

\begin{figure}[ht]
    \centering
    \includegraphics[width=1\textwidth]{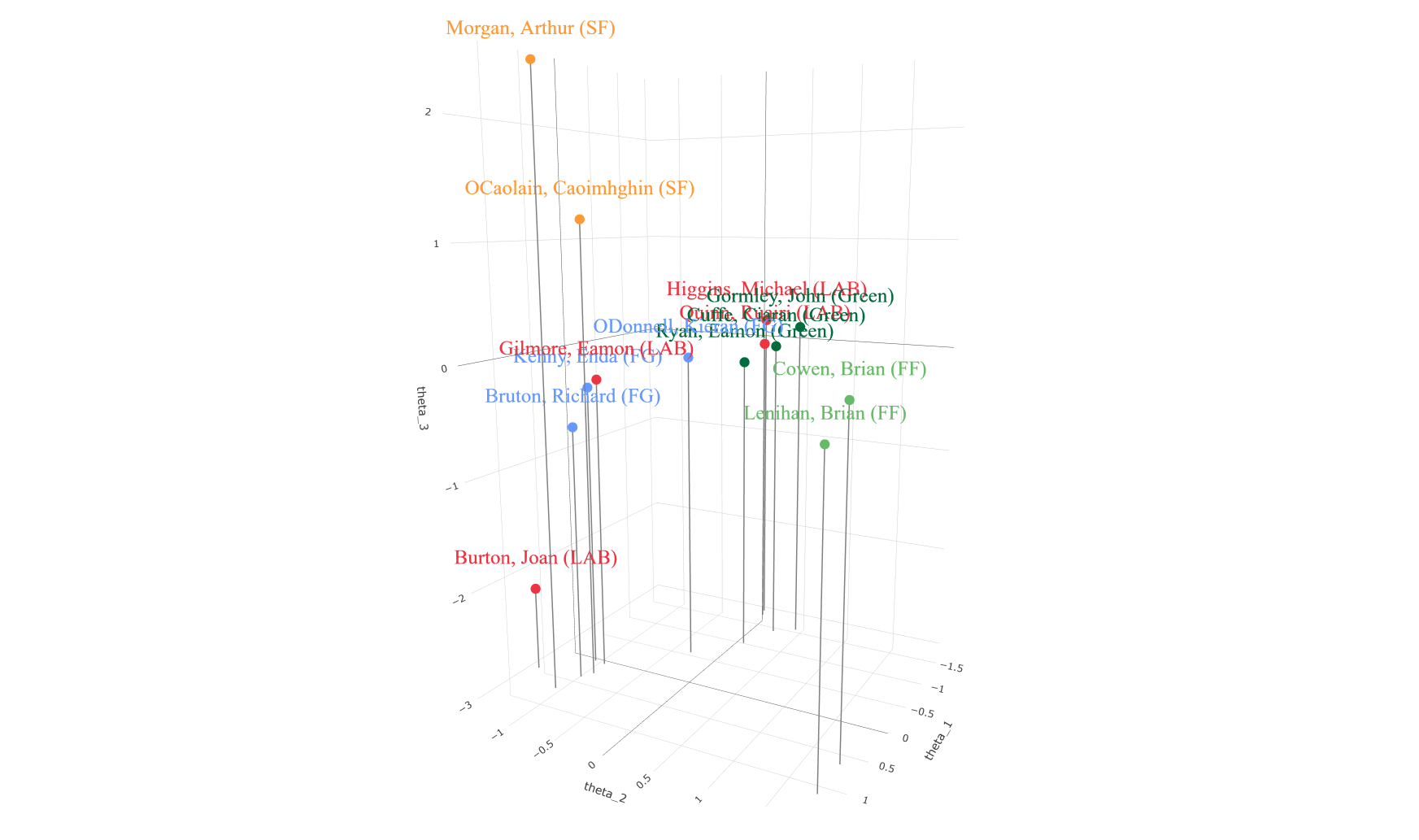}
    \caption{Position of Irish party members on three dimensions based on speeches in the Irish Dáil}
    \label{fig:scatter_IRL_wk3}
\end{figure}

Extending the Wordkrill model to three dimensions allows for a more nuanced understanding of MP positioning within the Irish Dáil. The resulting configuration, shown in Figure \ref{fig:scatter_IRL_wk3}, indicates that the clusters identified in the two-dimensional model remain relatively stable, while the third dimension primarily differentiates between Sinn Féin and Labour MPs within a previously shared cluster. This distinction aligns with the political context of the period: although both parties opposed the Fianna Fáil-led budget, they articulated their dissent through distinct ideological and rhetorical lenses. Joan Burton of the Labour Party framed her critique in terms of fiscal responsibility, institutional reform, and targeted social investment. In contrast, Sinn Féin foregrounded anti-austerity themes, working-class advocacy, and a more radical redistributive agenda. 
These distinctions are further reflected in the corresponding $\beta$–$\psi$ plot (Figure \ref{fig:scatter_IRL_wk3_beta3} in the appendix), where terms such as \textit{government}, \textit{workers}, and \textit{system} dominate the positive axis, while \textit{credit}, \textit{taxpayer}, and \textit{opportunities} characterise the negative end. This suggests that MPs within the broader anti-technocratic cluster vary in their degrees of populism and systemic critique.

In sum, expanding the dimensionality of the positioning space reveals meaningful intra-oppositional variation, highlighting the heterogeneity within the anti-austerity bloc. Rather than presenting a unified rhetorical front, parties deploy divergent frames and communicative strategies even when aligned on key policy positions. These findings underscore the model’s utility in thematically constrained but discursively diverse settings and its robustness when applied to speeches of medium length.

\subsection{Speeches in the German \textit{Bundestag}}
\label{subsec:GER_parliament}
As a third application example, speeches from the German parliament, the \textit{Bundestag}, are analysed \citep{ParlEE2022fulldata}. The data contains nearly 3 million sentences from over 200,000 speeches of 2014 speakers from six German parties.\footnote{Namely, the left-wing party \textit{Die Linke}, the far-right AfD, the Christian-conservative CDU/CSU, the neoliberal FDP, the green party \textit{Bündnis 90/Die Grünen} and the social democratic SPD}. All speeches are extracted from transcripts of sessions and have been supplemented with information about the content of the speech, the speaker's party and information about the speaker. To test the method in a thematically narrow field, only speeches on the topic of migration from the years 2015 to 2019 were analysed.\footnote{In total, 797 speeches with feature counts between 68 and just under 1000 remain in the corpus. These speeches fall under the overarching topic of migration, but in parliamentary debates, they focus on the sub-topic under discussion, e.g. asylum law and integration into the labour market. Whether Wordkrill can deal with this variation in thematic focus and the comparatively short speeches will be answered in this application example.} 
This topic and this period are an exciting example of application, as the so-called refugee crisis in 2015 produced a large number of speeches on migration \citep{geese_immigration-related_2020}, which were mostly positive and pragmatic before the far-right AfD entered the \textit{Bundestag} \citep{rakers_2015_2023}. With the entry of the right-wing AfD in 2017, the German party elites shifted on the socio-cultural dimension to the right \citep{jankowski_adapt_2019}. 
It would be expected that a successful application of Wordkrill to the above-mentioned speeches would detect meaningful underlying dimensions, identify clusters of MPs by party and show a shift in positioning after the entry of the AfD. 

\begin{figure}[ht]
    \centering
    \includegraphics[width=1\textwidth]{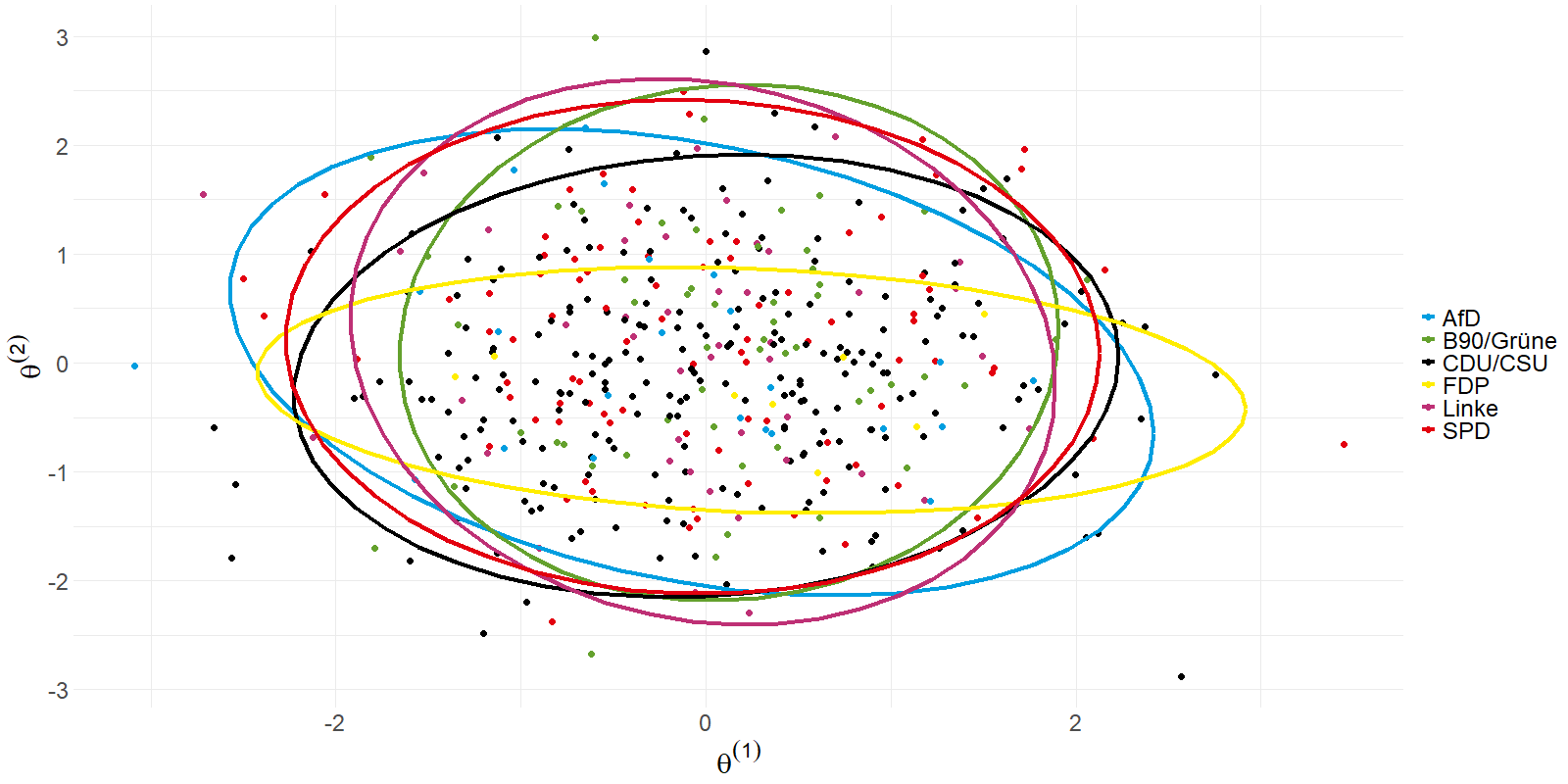}
    \caption{Position of MPs in the German \textit{Bundestag} on two dimensions based on speeches on migration}
    \label{fig:scatter_GERBT_wk2}
\end{figure}

\begin{figure}[ht]
    \centering
    \includegraphics[width=1\textwidth]{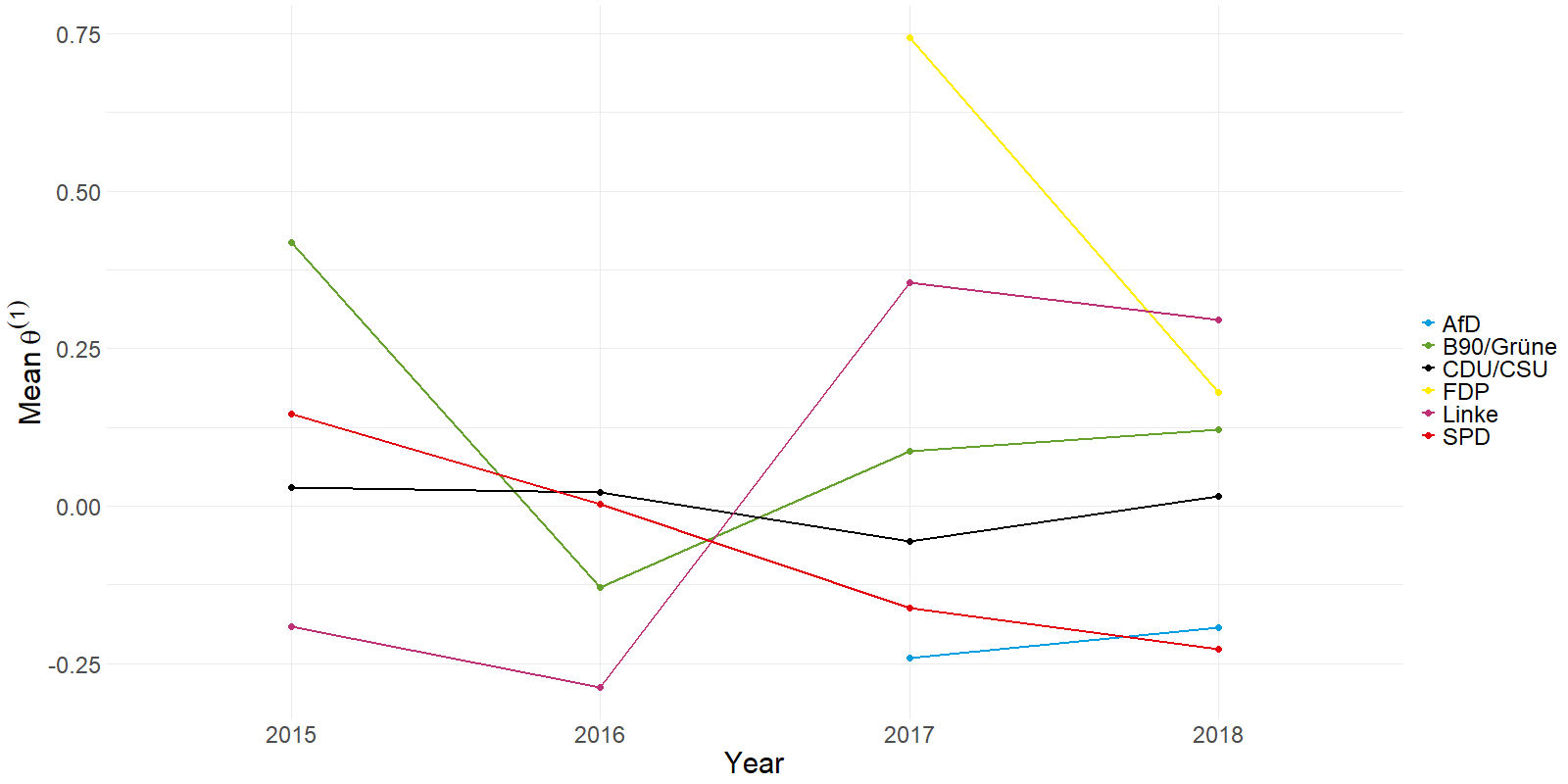}
    \caption{Mean position of MPs in the German \textit{Bundestag} on the first estimated dimension}
    \label{fig:line_GERBT_wk2_dim1}
\end{figure}

\begin{figure}[ht]
    \centering
    \includegraphics[width=1\textwidth]{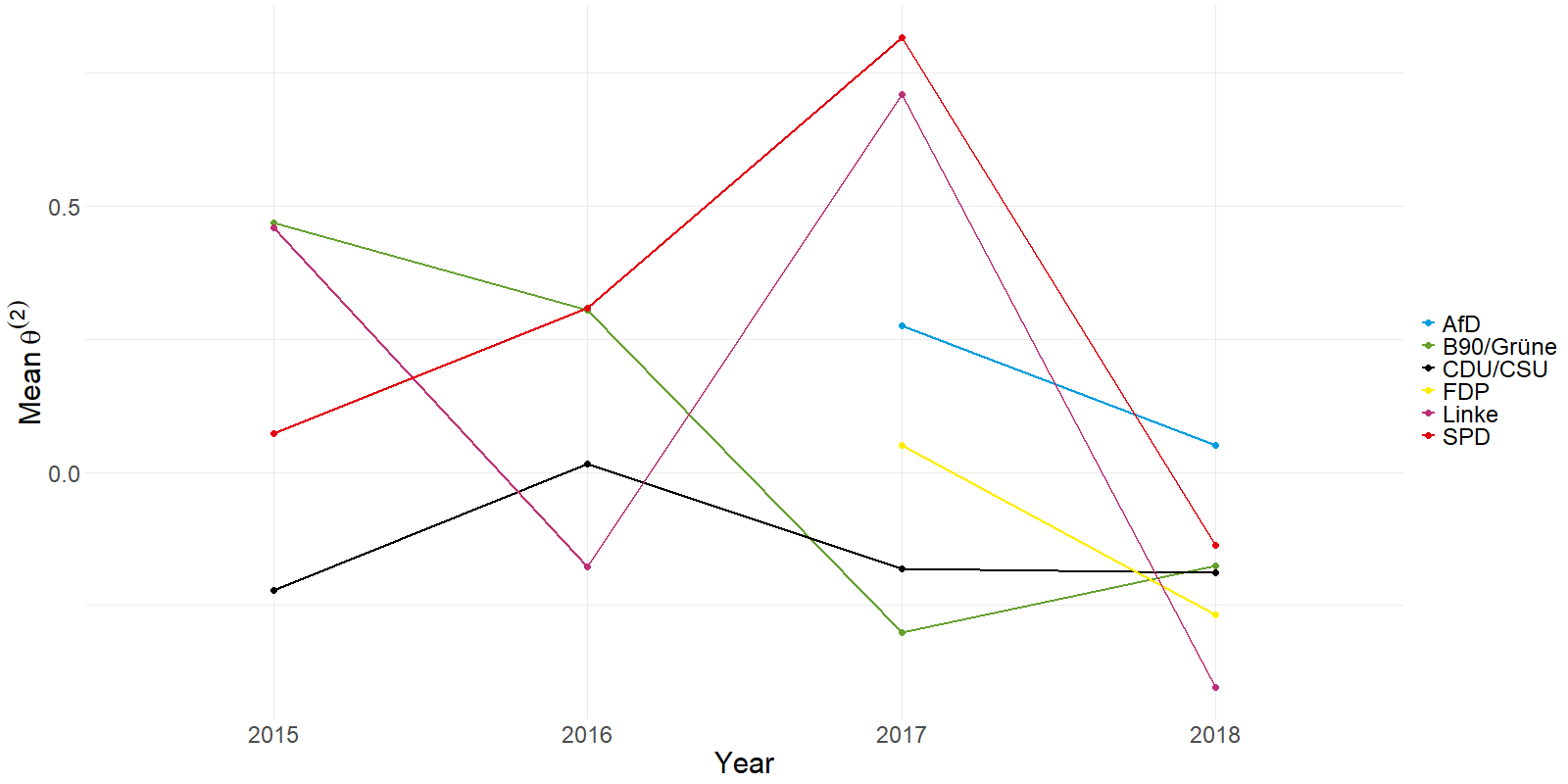}
    \caption{Mean position of MPs in the German \textit{Bundestag} on the second estimated dimension}
    \label{fig:line_GERBT_wk2_dim2}
\end{figure}

Figure \ref{fig:scatter_GERBT_wk2} shows the distribution of MPs on the two dimensions estimated by Wordkrill. The ellipses indicate the area in which the majority of MPs are located by party. Overall, it can be seen that the parties are not clustered, but rather chaotically distributed throughout the entire space. Even when looking at the average party positions over time in Figures \ref{fig:line_GERBT_wk2_dim1} and \ref{fig:line_GERBT_wk2_dim2}, there are neither clear divisions between the parties nor did the established parties appear to move uniformly in one direction after the entry of the AfD into parliament, indicating a shift in discourse.

To determine the meaning of the two dimensions, consider the $\beta$-$\psi$ plots in the appendix, see Figures \ref{fig:scatter_GERBT_wk2_beta1} and \ref{fig:scatter_GERBT_wk2_beta2}. There is no clear distinction here between critical/disparaging and supportive/enhancing features: For example, \textit{hospitality} is found on the positive side for the first dimension, while \textit{helpful} is found on the negative side. Incongruent features such as \textit{security situation} and \textit{scaremongering} can also be found on the same side. 
The same applies to the second dimension, where no clear ordering of the features is recognisable either. However, it is noticeable that the second dimension is dominated by organisational features, such as \textit{deportation ban} or \textit{decider}, and less by symbolic, radical features. 
To summarise, it can be seen that the proposed model is not suitable for analysing the migration discourse in Germany. This is certainly due to the sometimes quite short speeches of less than 100 features, but also due to the varying thematic focus (when discussing individual, specific legislative initiatives).\footnote{Increasing or reducing the dimensions $K$ does not lead to valid results either. In the three-dimensional case, the third dimension is mainly shaped by time effects. With one dimension, the positioning also remains chaotic and the meaning of the dimension unclear.}

\section{Conclusion and Outlook}

This paper introduced Wordkrill, a multidimensional extension of the widely used Wordfish model, to better capture the inherently complex and multidimensional nature of politics and political text data \citep{gherghina2019dynamics}. Wordkrill addresses a critical limitation of the original model and offers a more nuanced, data-driven representation of political discourse by allowing for the estimation of document positions along multiple latent dimensions. 

The empirical applications demonstrate that Wordkrill performs well in contexts where political competition is structured along clear ideological axes, such as party manifestos and delimited parliamentary debates. In these cases, the estimated dimensions align with established theoretical constructs like the socio-economic and socio-cultural cleavage or technocratic vs. populist rhetoric, offering both interpretive clarity and empirical validity. However, the model proves less effective in cases with fragmented discourse, short texts, or diffuse thematic focus, as illustrated by the analysis of \textit{Bundestag} speeches on migration. This highlights the importance of context and corpus characteristics when applying automated text scaling methods. The model also inherits the known disadvantages and limitations of the underlying Wordfish model.

Despite these limitations, Wordkrill represents a valuable methodological advancement. Its unsupervised nature, scalability, and interpretability make it particularly attractive for analysing large corpora where supervised methods are infeasible or costly. Moreover, Wordkrill brings researchers closer to how political actors position themselves, through the very language they use, without imposing a priori dimensional structures. 
Future work may refine the model’s estimation techniques, explore non-linear or hierarchical extensions, and test its performance across languages, media, and institutional contexts. 
In sum, Wordkrill expands the methodological repertoire for political text analysis and opens new avenues for the empirical study of multidimensional political competition

\newpage
\FloatBarrier


\bibliography{references}

\appendix
\section{Appendix}

\subsection{German Party Manifestos}
\begin{figure}[ht]
    \centering
    \includegraphics[width=1\textwidth]{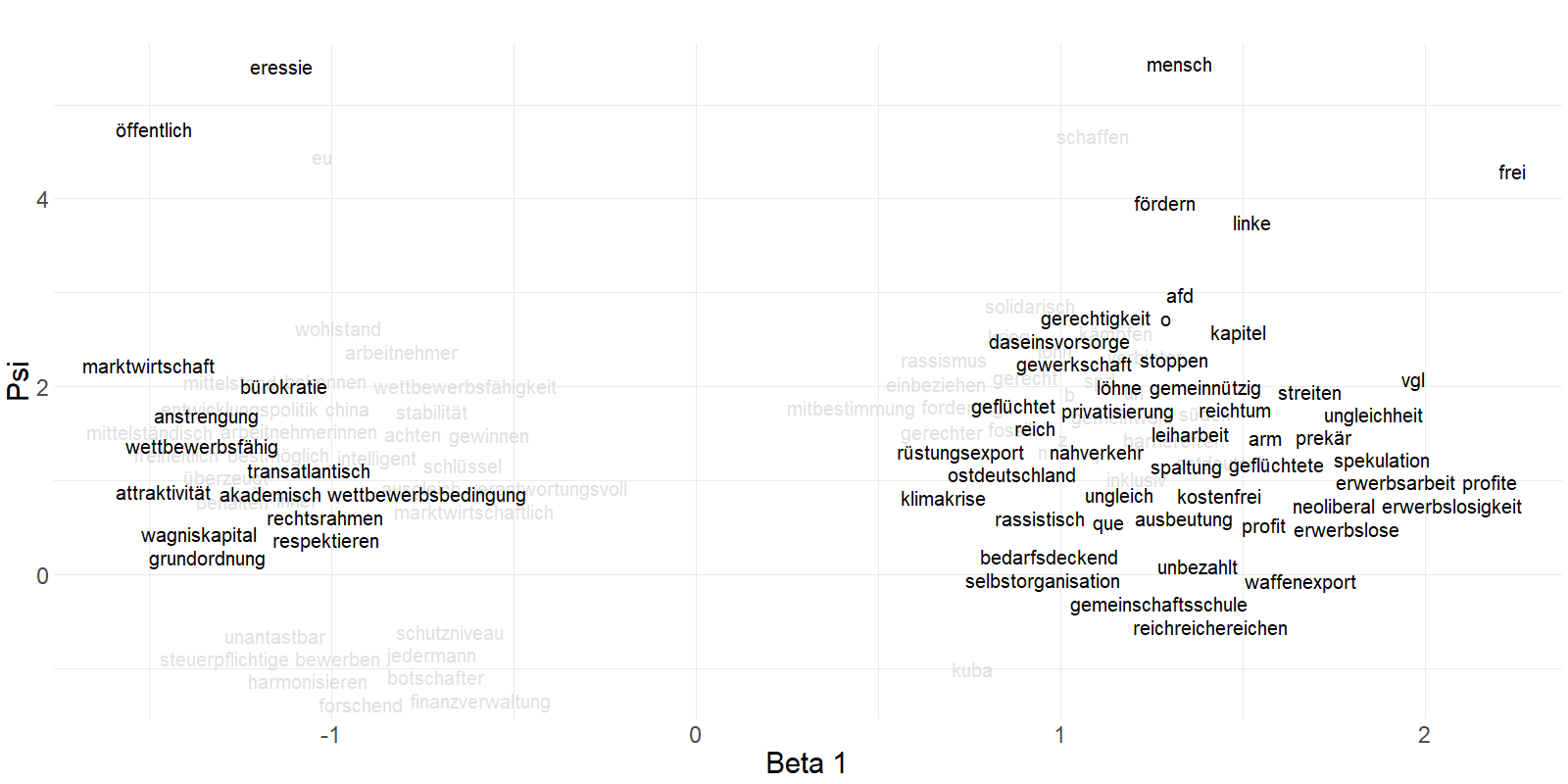}
    \caption{Distribution of $\beta^{(1)}$ vs. $\psi$ values for German party manifestos, Wordkrill with two dimensions}
    \label{fig:scatter_GER_wk2_beta1}
\end{figure}

\begin{figure}[ht]
    \centering
    \includegraphics[width=1\textwidth]{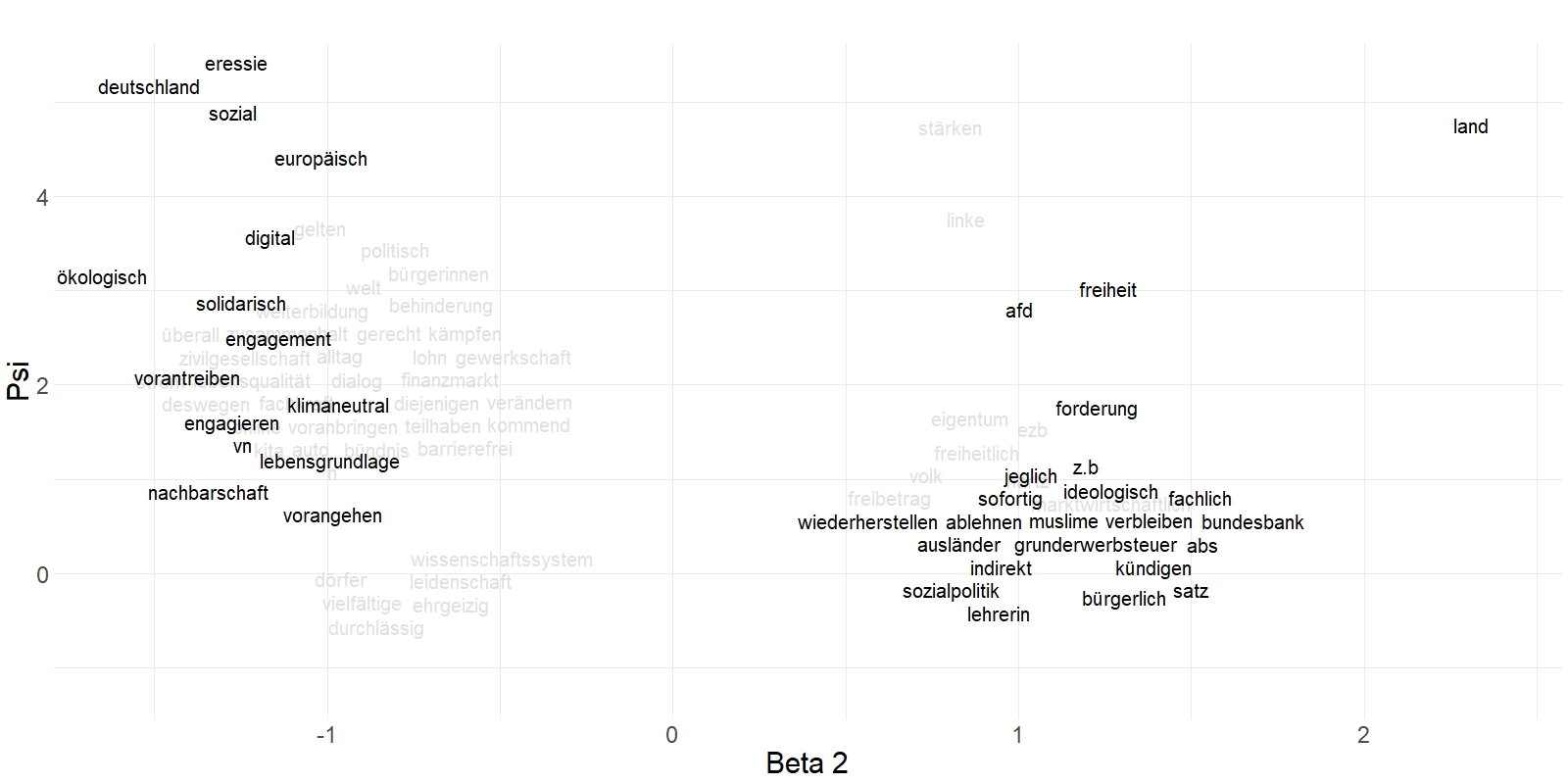}
    \caption{Distribution of $\beta^{(2)}$ vs. $\psi$ values for German party manifestos, Wordkrill with two dimensions}
    \label{fig:scatter_GER_wk2_beta2}
\end{figure}

\begin{figure}[ht]
    \centering
    \includegraphics[width=1\textwidth]{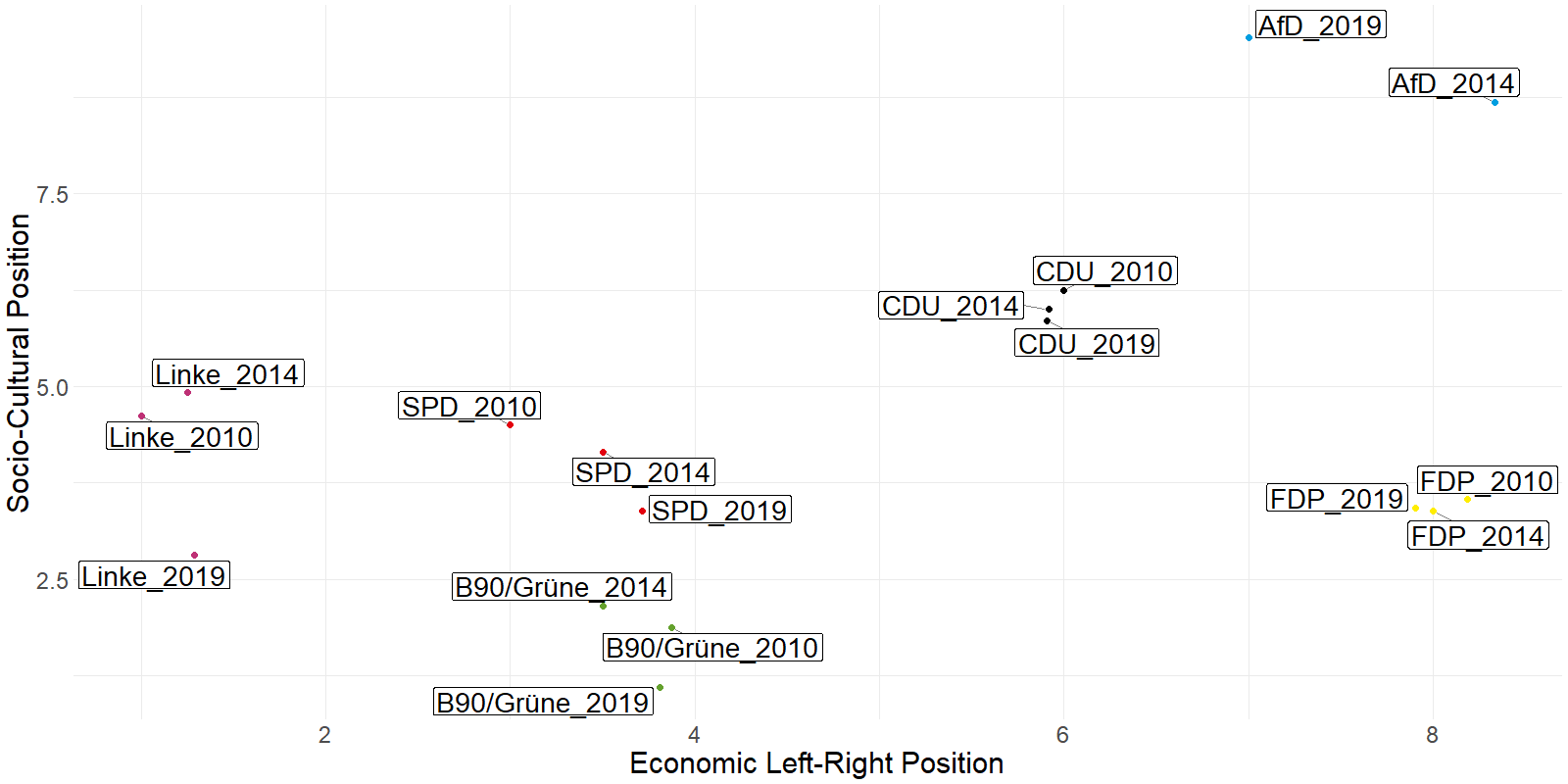}
    \caption{Position of the German parties on the socio-cultural and socio-economic dimensions \citep{jolly2022chapel}}
    \label{fig:scatter_GER_CHES_SESC}
\end{figure}

\begin{figure}[ht]
    \centering
    \includegraphics[width=1\textwidth]{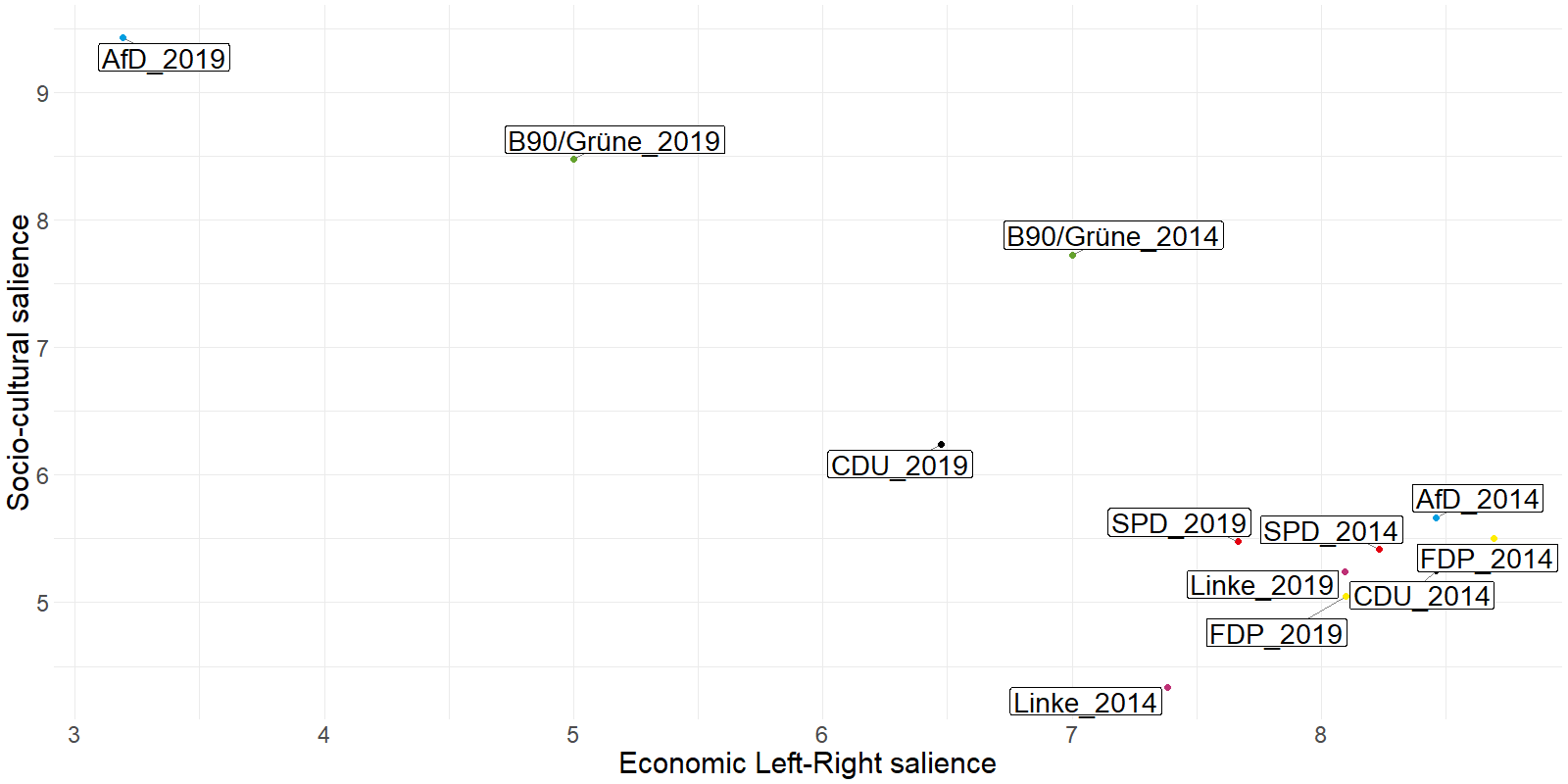}
    \caption{Salience of the socio-cultural and socio-economic dimensions for the German parties}
    \label{fig:scatter_GER_CHES_SESC_SAL}
\end{figure}

\clearpage
\FloatBarrier
\subsection{Speeches in the Irish Dáil}
\begin{figure}[ht]
    \centering
    \includegraphics[width=1\textwidth]{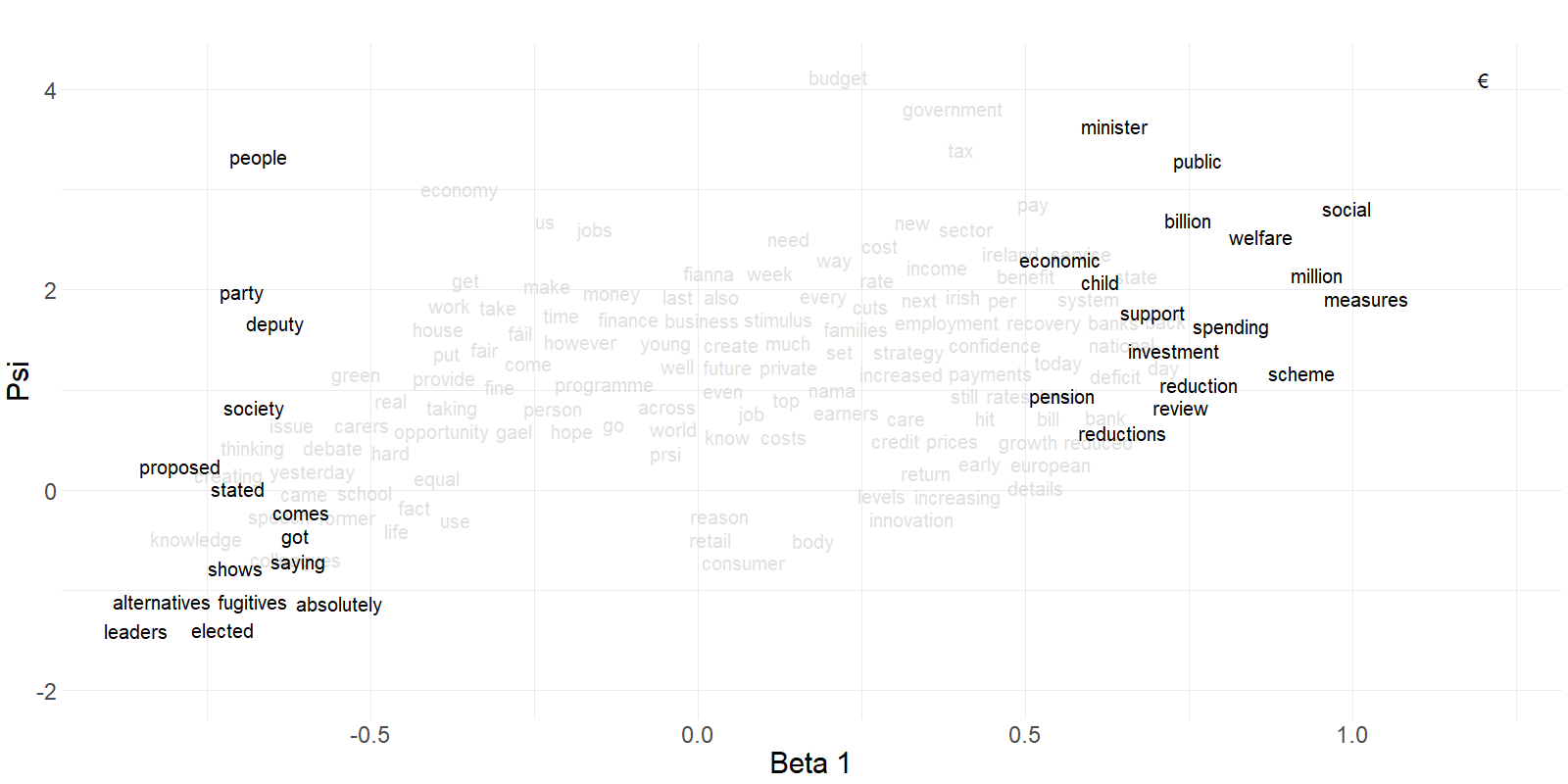}
    \caption{Distribution of $\beta^{(1)}$ vs. $\psi$ values for speeches in the Irish Dáil, Wordkrill with two dimensions}
    \label{fig:scatter_IRL_wk2_beta1}
\end{figure}

\begin{figure}[ht]
    \centering
    \includegraphics[width=1\textwidth]{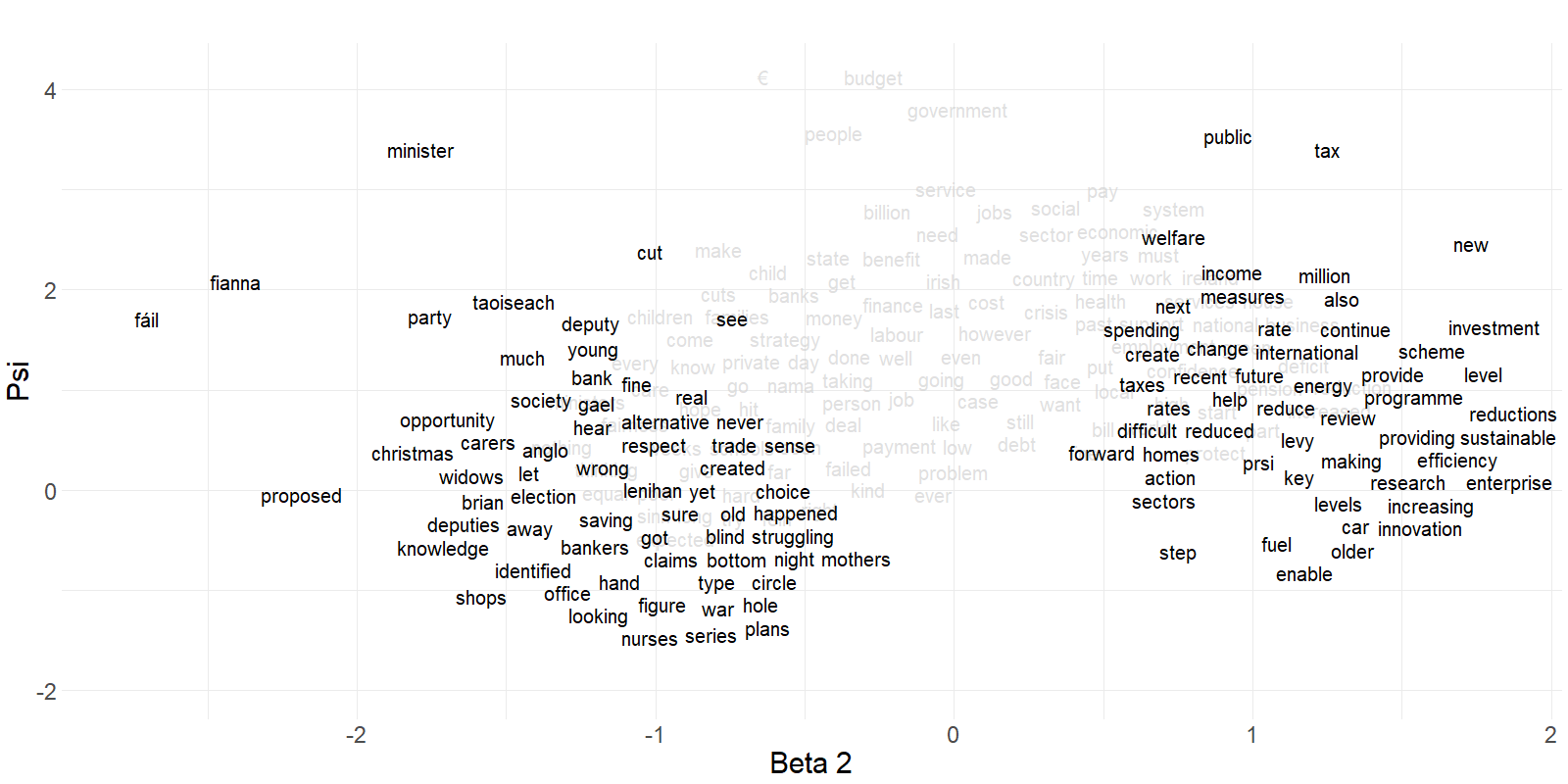}
    \caption{Distribution of $\beta^{(2)}$ vs. $\psi$ values for speeches in the Irish Dáil, Wordkrill with two dimensions}
    \label{fig:scatter_IRL_wk2_beta2}
\end{figure}

\begin{figure}[ht]
    \centering
    \includegraphics[width=1\textwidth]{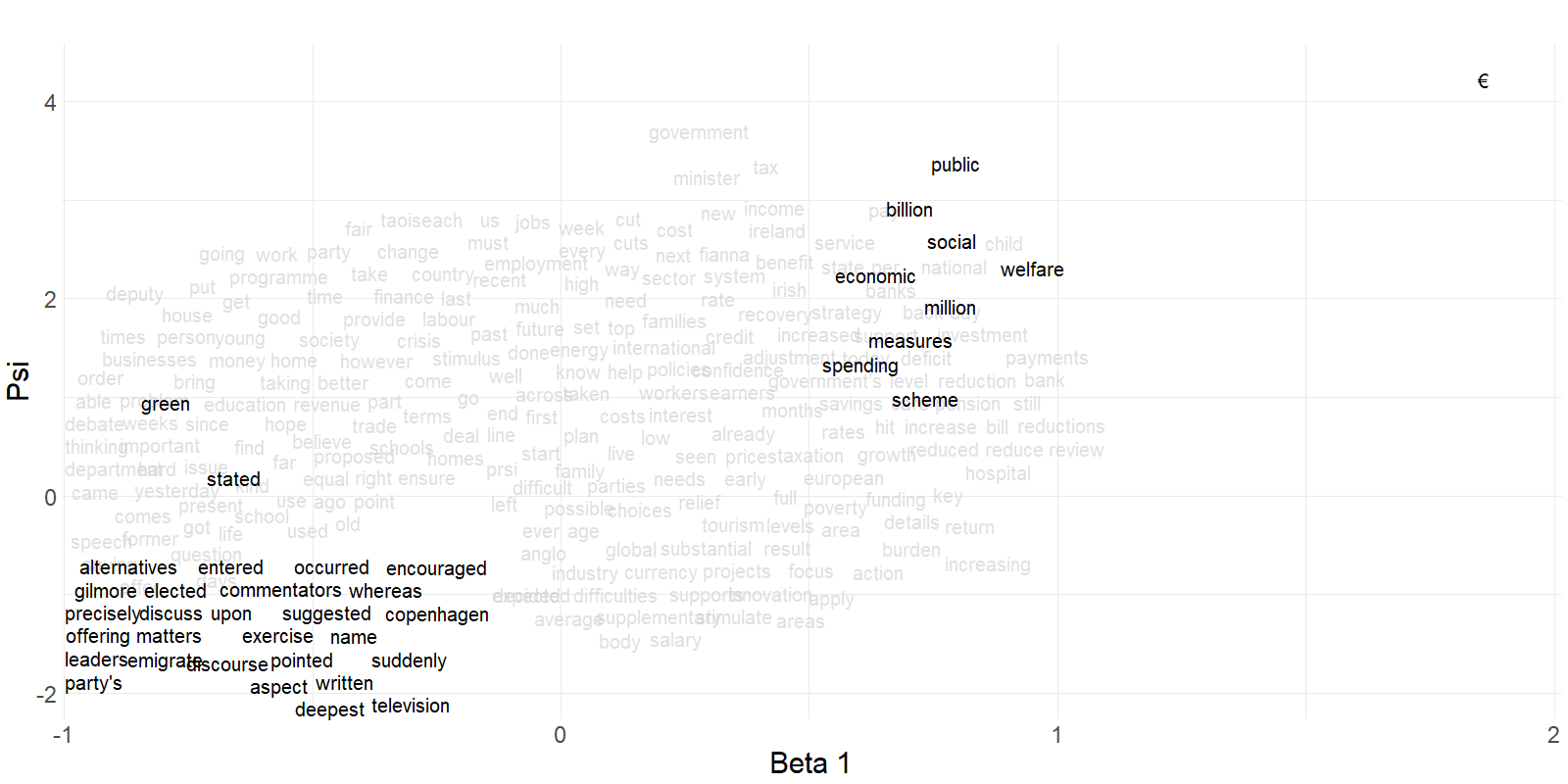}
    \caption{Distribution of $\beta^{(1)}$ vs. $\psi$ values for speeches in the Irish Dáil, Wordkrill with three dimensions}
    \label{fig:scatter_IRL_wk3_beta1}
\end{figure}

\begin{figure}[ht]
    \centering
    \includegraphics[width=1\textwidth]{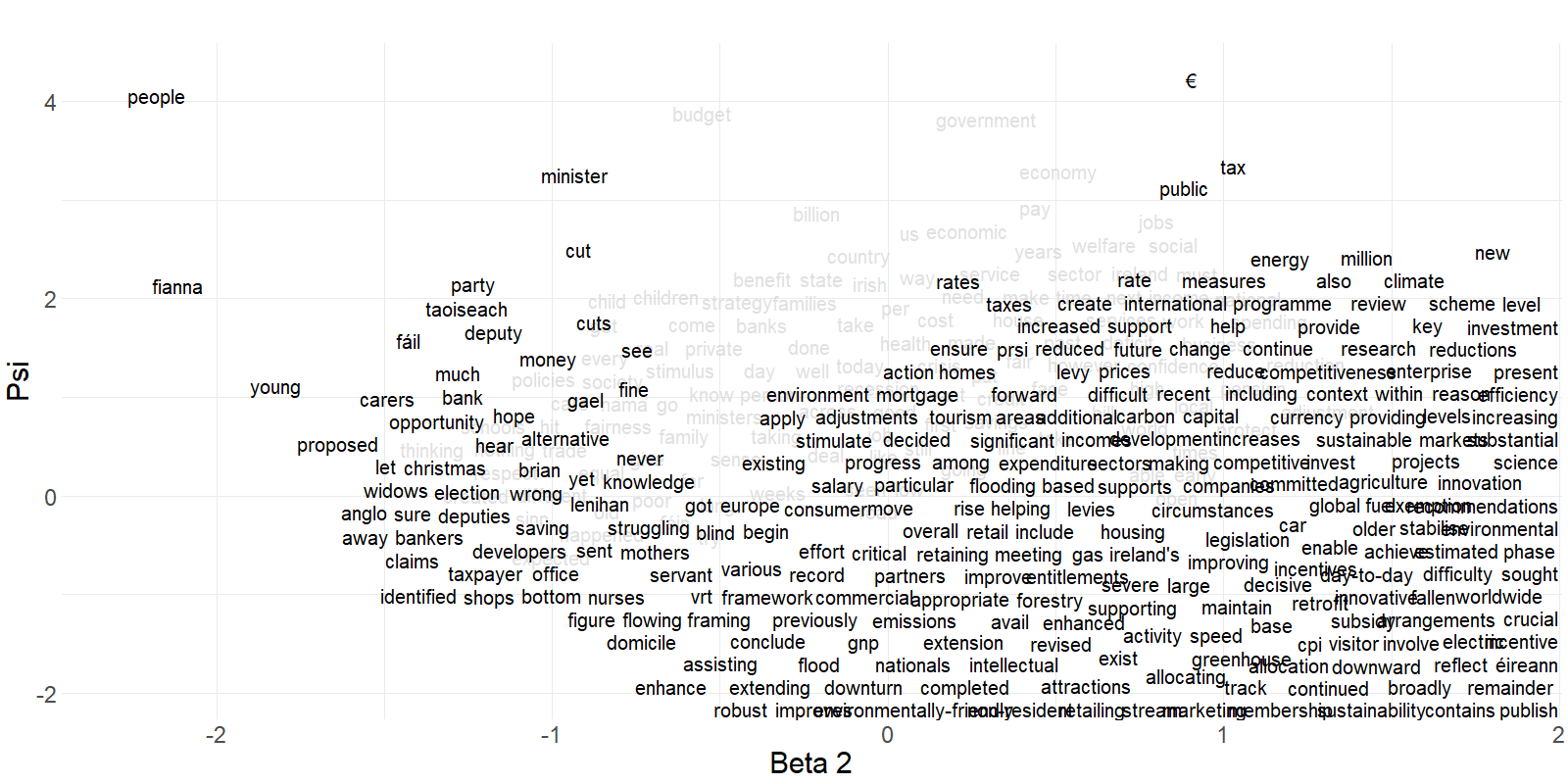}
    \caption{Distribution of $\beta^{(2)}$ vs. $\psi$ values for speeches in the Irish Dáil, Wordkrill with three dimensions}
    \label{fig:scatter_IRL_wk3_beta2}
\end{figure}

\begin{figure}[ht]
    \centering
    \includegraphics[width=1\textwidth]{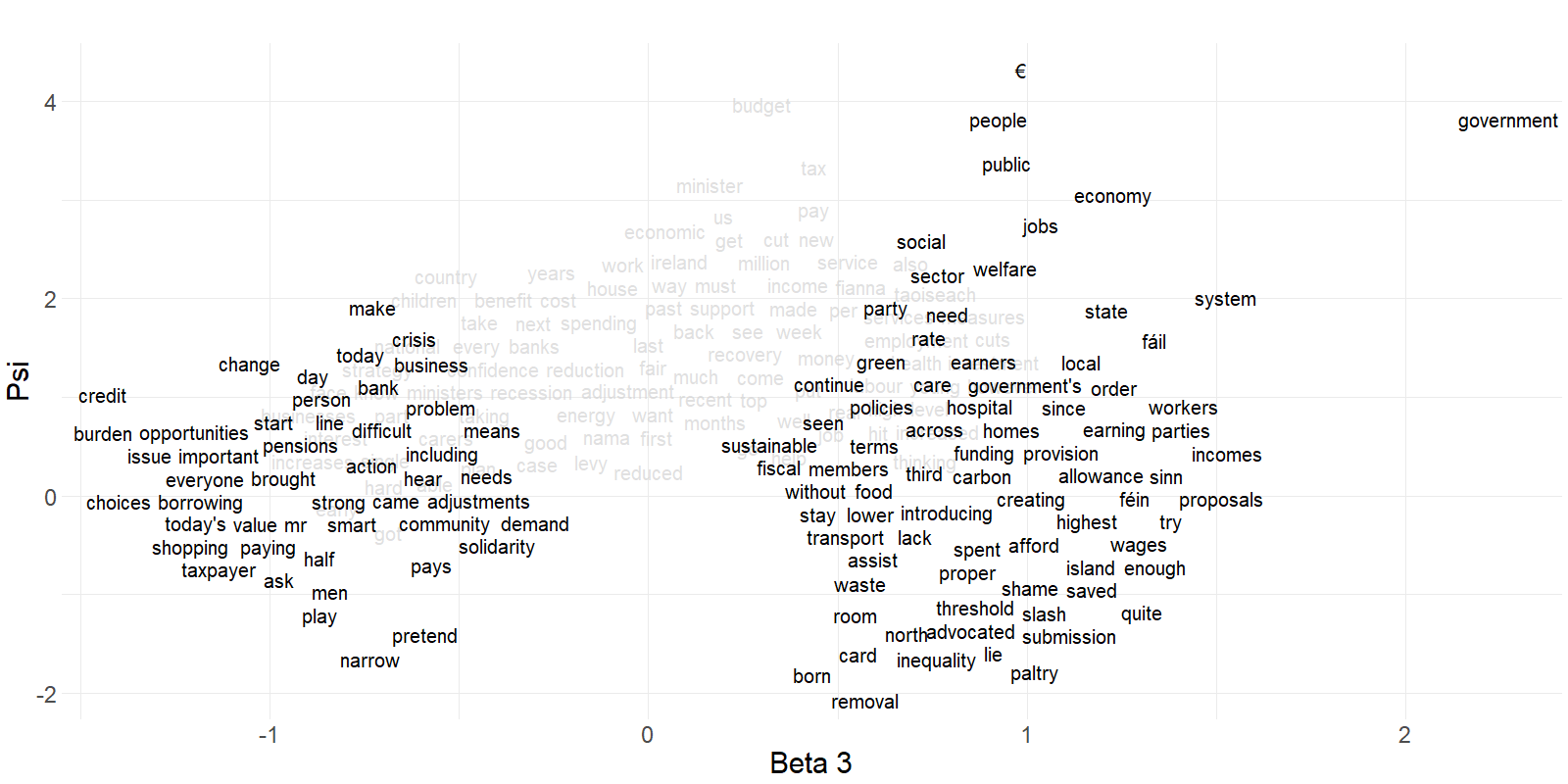}
    \caption{Distribution of $\beta^{(3)}$ vs. $\psi$ values for speeches in the Irish Dáil, Wordkrill with three dimensions}
    \label{fig:scatter_IRL_wk3_beta3}
\end{figure}

\FloatBarrier
\subsection{Speeches in the German \textit{Bundestag}}

\begin{figure}[ht]
    \centering
    \includegraphics[width=1\textwidth]{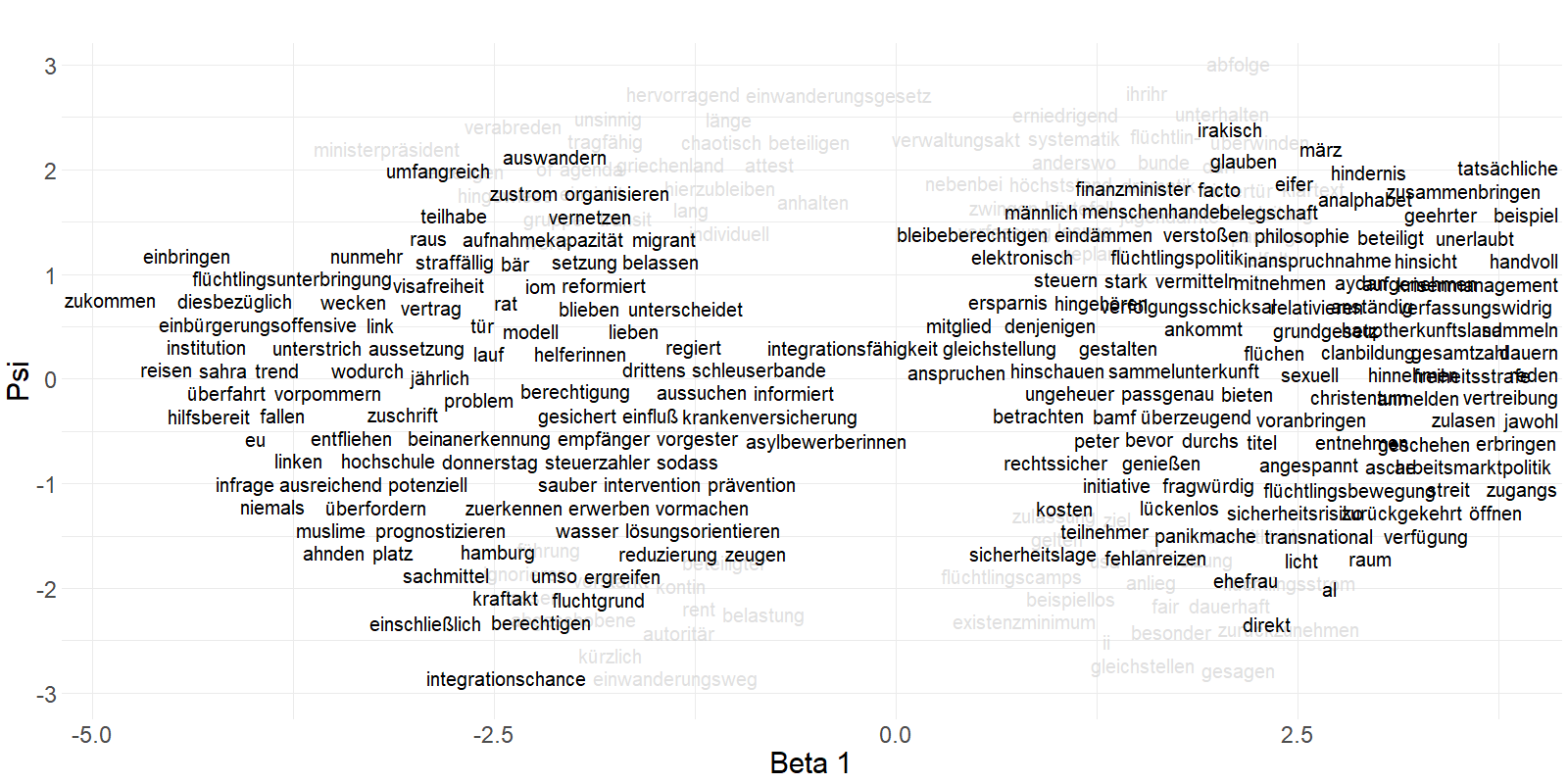}
    \caption{Distribution of $\beta^{(1)}$ vs. $\psi$ values for speeches in the German \textit{Bundestag}, Wordkrill with two dimensions}
    \label{fig:scatter_GERBT_wk2_beta1}
\end{figure}

\begin{figure}[ht]
    \centering
    \includegraphics[width=1\textwidth]{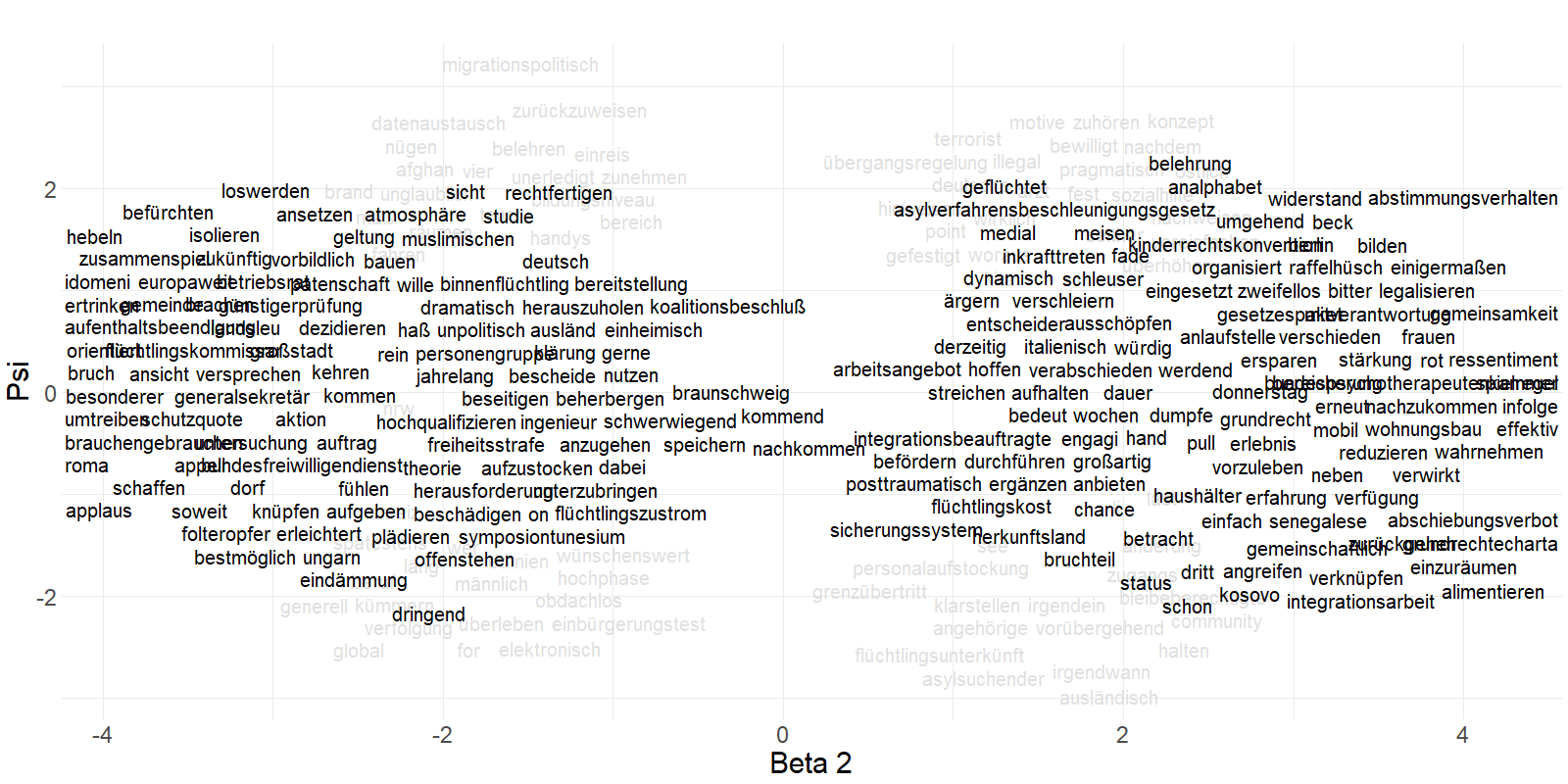}
    \caption{Distribution of $\beta^{(2)}$ vs. $\psi$ values for speeches in the German \textit{Bundestag}, Wordkrill with two dimensions}
    \label{fig:scatter_GERBT_wk2_beta2}
\end{figure}

\end{document}